\documentclass[letterpaper,twocolumn,10pt]{article}
\usepackage{usenix2019_v3}

\usepackage{setspace}
\usepackage{listings}
\usepackage{amsfonts}
\usepackage{graphicx}
\usepackage{lstlinebgrd}
\usepackage[ruled,vlined,linesnumbered,nofillcomment,scleft]{algorithm2e}

\SetCommentSty{mycommfont}
\usepackage[shortlabels]{enumitem}
\setlist[enumerate]{
    noitemsep,
    leftmargin=*,
    topsep=0pt,
    parsep=0pt,
    labelindent=0pt
}

\usepackage{xspace}
\usepackage{tcolorbox}
\usepackage{tikz}
\usepackage{makecell}
\usepackage{booktabs}
\usepackage{colortbl}

\newcolumntype{B}{>{\columncolor[rgb]{0.85,.75,.75}}}
\newcolumntype{C}{>{\columncolor[rgb]{0.75,.85,.75}}}
\newcolumntype{A}{>{\columncolor[rgb]{0.97,.97,.97}}}
\newcolumntype{D}{>{\columncolor[rgb]{0.94,.94,.94}}}
\newcolumntype{E}{>{\columncolor[rgb]{0.91,.91,.91}}}

\usepackage{caption}
\usepackage[font=footnotesize]{subfig}

\usepackage{multirow}
\usepackage{pifont}

\usepackage{array}

\newcolumntype{?}[1]{!{\color{white}\vrule width #1 }}

\newcommand*{\yyy}{\ding{51}}
\newcommand*{\xxx}{\ding{55}}

\newcommand*{\testsALL}{\ensuremath{N_{{all}}}\xspace}
\newcommand*{\testsMT}{\ensuremath{N_{{mt}}}\xspace}
\newcommand*{\testsRatio}{$\frac{\testsMT}{\testsALL}$\xspace}
\newcommand*{\VulsMT}{\ensuremath{V_{m}}\xspace}
\newcommand*{\NVulsMT}{\ensuremath{N_{v}^m}\xspace}
\newcommand*{\VulsST}{\ensuremath{V_{s}}\xspace}
\newcommand*{\NVulsST}{\ensuremath{N_{v}^s}\xspace}
\newcommand*{\VulsCB}{\ensuremath{V_{cb}}\xspace}

\newcommand*{\BugsMT}{\ensuremath{B_{m}}\xspace}
\newcommand*{\crashesNUM}{\ensuremath{N_{c}}\xspace}
\newcommand*{\crashesMT}{\ensuremath{N_{c}^m}\xspace}
\newcommand*{\crashesST}{\ensuremath{N_{c}^s}\xspace}

\newcommand*{\codefont}{\texttt}
\newcommand*{\algofont}{\textit}
\newcommand*{\projfont}{\emph}

\newcommand*{\ccr}[1]{\ensuremath{P_e(#1)}\xspace}

\newcommand*{\mtp}[2]{{\ensuremath{P_m(#1, #2)}}\xspace}

\newcommand*{\mtpnO}{{\ensuremath{P_{m0}}}\xspace}

\newcommand*{\stp}[1]{{\ensuremath{P_s(#1)}}\xspace}
\newcommand*{\stpn}{{\ensuremath{P_s}}\xspace}
\newcommand*{\stpnO}{{\ensuremath{P_{s0}}}\xspace}

\newcommand*{\rtifunc}{\ensuremath{{F}_{S}}\xspace}

\newcommand*{\aBB}{\ensuremath{b}\xspace}
\newcommand*{\aFUNC}{\ensuremath{f}\xspace}
\newcommand*{\aINSTR}{\ensuremath{i}\xspace}

\newcommand*{\ntid}{\ensuremath{N_{ctx}}\xspace}
\newcommand*{\iloc}{\ensuremath{Loc}\xspace}
\newcommand*{\threadCtx}[2]{\ensuremath{\langle #1, #2\rangle}\xspace}
\newcommand*{\FSThread}{\ensuremath{F_{ctx}}\xspace}
\newcommand*{\threadCtxn}{\ensuremath{TC}}

\newcommand*{\tctxSign}{\ensuremath{{S}_{ctx}}\xspace}

\newcommand*{\FSStart}{\ensuremath{F_{fork}}\xspace}

\newcommand*{\mutChance}{\ensuremath{\mathbb{M}}\xspace}

\newcommand*{\probNNN}{\ensuremath{P_{nnn}}\xspace}
\newcommand*{\probYNT}{\ensuremath{P_{ynt}}\xspace}
\newcommand*{\probYNN}{\ensuremath{P_{ynn}}\xspace}

\newcommand*{\NcalO}{\ensuremath{N_0}\xspace}
\newcommand*{\NcalV}{\ensuremath{N_v}\xspace}
\newcommand*{\NcalB}{\ensuremath{B_v}}
\newcommand*{\Ncal}{\ensuremath{\mathbb{N}_{c}}\xspace}

\newcommand*{\Nmt}{\ensuremath{C_{m}}\xspace}

\newcommand*{\Seeds}{\ensuremath{Q_{S}}\xspace}
\newcommand*{\FinalSeeds}{\ensuremath{Q_{S}}\xspace}
\newcommand*{\CrashSeeds}{\ensuremath{T_{C}}\xspace}
\newcommand*{\Prog}{\ensuremath{\mathbb{P}_f}\xspace}
\newcommand*{\ProgO}{\ensuremath{\mathbb{P}_o}\xspace}

\tikzset{nodescale/.style = {scale=#1}}
\newcommand\encircle[1]{
  \tikz[baseline={([yshift=-.2ex]X.base)}] 
    \node (X) [draw, shape=circle, inner sep=0, nodescale={0.8}] {\strut{#1}};}
\newcommand*{\PinstS}{\encircle{\small{A}}\xspace}
\newcommand*{\PfuzzS}{\encircle{\small{B}}\xspace}
\newcommand*{\PvulS}{\encircle{\small{C}}\xspace}
\newcommand*{\PcbugS}{\encircle{\small{D}}\xspace}
\newcommand*{\PinstF}{\encircle{\scriptsize{A}}\xspace}
\newcommand*{\PfuzzF}{\encircle{\scriptsize{B}}\xspace}
\newcommand*{\PvulF}{\encircle{\scriptsize{C}}\xspace}
\newcommand*{\PcbugF}{\encircle{\scriptsize{D}}\xspace}
\newcommand\circleNUM[1]{
  \tikz[baseline={([yshift=-.6ex]current bounding box.center)}]
    \node (X) [draw, shape=circle, inner sep=0, nodescale={0.7}] {\strut{\footnotesize{#1}}};}
\newcommand\circleNUMT[1]{
  \tikz[baseline={([yshift=-.25ex]X.base)}] 
    \node (X) [draw, shape=circle, inner sep=0, nodescale={0.7}] {\strut{\footnotesize{#1}}};}

\newcommand*{\ts}{{\textmd{TSan}}\xspace}

\newcommand*{\muzz}{{\textsc{Muzz}}\xspace}
\newcommand*{\mafl}{\textsc{MAFL}\xspace}
\newcommand*{\mopt}{\textsc{MOpt}\xspace}

\newcommand*{\mtiscope}{\ensuremath{L_{m}}\xspace}

\newcommand*{\mtcolor}{\color{lightgray}}

\newcommand*{\replayRO}{\ensuremath{\mathbb{P}1}\xspace}
\newcommand*{\replayRN}{\ensuremath{\mathbb{P}2}\xspace}
\newcommand*{\replayCN}{\ensuremath{N_{e}^m}}
\newcommand*{\replayCB}{\ensuremath{N_{B}^m}}

\newcommand*{\pbin}[1]{\textcolor{black}{#1}}
\newcommand*{\gres}[1]{\textbf{#1}}
\newcommand*{\mydiffn}[1]{}
\newcommand*{\bres}[1]{\underline{#1}}
\newcommand*{\mydiff}[1]{(#1)}

\newcommand*{\AFLIns}{\textsf{AFL-Ins}\xspace}
\newcommand*{\MTIns}{\textsf{M-Ins}\xspace}

\newcommand*{\TStart}{\algofont{TFork}\xspace}
\newcommand*{\TEnd}{\algofont{TJoin}\xspace}
\newcommand*{\TLock}{\algofont{TLock}\xspace}
\newcommand*{\TUnLock}{\algofont{TUnLock}\xspace}
\newcommand*{\TSharedVar}{\algofont{TShareVar}\xspace}

\begin{document}

\date{}
\title{\Large \bf \muzz: Thread-aware Grey-box Fuzzing \\
for Effective Bug Hunting in Multithreaded Programs}

\author{\rm Hongxu Chen\textsuperscript{\S}\textsuperscript{\dag} \quad Shengjian Guo\textsuperscript{\ddag} \quad Yinxing Xue\textsuperscript{\S}\thanks{Corresponding Author.} \quad Yulei Sui\textsuperscript{\P} \\
\rm Cen Zhang\textsuperscript{\dag}\quad Yuekang Li\textsuperscript{\dag} \quad Haijun Wang\textsuperscript{\#} \quad Yang Liu\textsuperscript{\dag} \\
\textsuperscript{\dag}Nanyang Technological University \qquad \textsuperscript{\ddag}Baidu Security \qquad
\textsuperscript{\P}University of Technology Sydney \\
\textsuperscript{\S}University of Science and Technology of China \qquad
\textsuperscript{\#}Ant Financial Services Group
}
\maketitle

\thispagestyle{plain}
\pagestyle{plain}

\begin{abstract}
Grey-box fuzz testing has revealed thousands of vulnerabilities
in real-world software owing to its lightweight instrumentation, fast 
coverage feedback, and dynamic adjusting strategies. However, directly 
applying grey-box fuzzing to input-dependent multithreaded programs can 
be extremely inefficient. In practice, multithreading-relevant bugs are usually buried in the sophisticated program flows.
Meanwhile, existing grey-box fuzzing techniques do not stress thread-interleavings that affect execution states in multithreaded programs.
Therefore, mainstream grey-box fuzzers cannot adequately test problematic segments in
multithreaded software, although they might obtain high code coverage 
statistics.

To this end, we propose \muzz, a new grey-box fuzzing technique that hunts 
for bugs in multithreaded programs. \muzz owns three novel thread-aware 
instrumentations, namely coverage-oriented instrumentation, thread-context instrumentation, and schedule-intervention instrumentation. During fuzzing, 
these instrumentations engender runtime feedback to accentuate execution states 
caused by thread interleavings. By leveraging such feedback in the dynamic 
seed selection and execution strategies, \muzz preserves more valuable 
seeds that expose bugs under a multithreading context.

We evaluate \muzz on twelve real-world multithreaded programs. Experiments show 
that \muzz outperforms AFL in both multithreading-relevant seed generation 
and concurrency-vulnerability detection. Further, by replaying the target 
programs against the generated seeds, \muzz also reveals more concurrency-bugs
(e.g., data-races, thread-leaks) than AFL. In total, \muzz detected eight new 
concurrency-vulnerabilities and nineteen new concurrency-bugs. At the time of 
writing, four reported issues have received CVE IDs.

\end{abstract}

\section{Introduction}\label{sec:intro}

Multithreading has been popular in modern software systems since 
it substantially utilizes the hardware resources to boost software
performance.
A typical computing paradigm of multithreaded programs is to accept a set of inputs, distribute computing jobs to threads, and orchestrate their progress accordingly. Compared to sequential programs, however, multithreaded programs 
are more prone to severe software faults. On the one hand, the non-deterministic
thread-interleavings give rise to \emph{concurrency-bugs} like data-races,
deadlocks, etc~\cite{cbugs_lu}. These bugs may cause the program to end up
with abnormal results or unexpected hangs. On the other hand, bugs that appear 
under specific inputs and interleavings may lead to \emph{concurrency-vulnerabilities}~\cite{ConAFL,CaiZMYHSL19}, resulting in memory corruptions, information leakage, etc.

There exist a line of works on detecting bugs and vulnerabilities in
multithreaded programs. Static concurrency-bug predictors
~\cite{pratikakis2006locksmith,Vojdani2009,SuiDX16,racerdoopsla2018}
aim to approximate the runtime behaviors of a program without actual concurrent execution. However, they typically serve as a complementary solution due to the high percentage of false alarms~\cite{mtbugs_survey}.
Dynamic detectors detect concurrency-violations by reasoning memory
read/write and synchronization events in a particular execution trace
\cite{YangCGK07,pldi09_fasttrack,lockset_SavageABNS97,kcc:tsan,helgrind,icse18_ufo,CaiZMYHSL19}. Several techniques like ThreadSanitizer (a.k.a., \ts)~\cite{kcc:tsan} and 
Helgrind~\cite{helgrind} have been widely used in practice. 
However, these approaches 
by themselves
do not automatically {\emph{generate
new test inputs} to exercise different paths in multithreaded programs}.

Meanwhile, grey-box fuzzing is effective in generating test inputs to expose vulnerabilities 
\cite{fuzzing1990,fuzz_survey}. It is reported that grey-box fuzzers 
(GBFs) such as AFL~\cite{afl_detail} and libFuzzer~\cite{libfuzzer} have detected
more than 16,000 vulnerabilities in hundreds of real-world software projects 
\cite{afl_detail,libfuzzer,clusterfuzz}.

Despite the great success of GBFs in detecting vulnerabilities, \emph{there are few efforts on fuzzing user-space
multithreaded programs}. General-purpose GBFs usually cannot explore thread-interleaving introduced execution states due to their unawareness of multithreading. Therefore, they cannot effectively detect concurrency-vulnerabilities inherently buried in sophisticated program flows~\cite{ConAFL}. In a discussion in 2015~\cite{afl-mt-st}, the author of AFL, Michal Zalewski, even suggests that ``\emph{it's generally better to have a single thread}''. In fact, due to the difficulty and inefficiency, the fuzzing driver programs in Google's continuous fuzzing platform OSS-fuzz are all tested in \emph{single-threaded} mode~\cite{oss-fuzz-proj}. Also, by matching unions of keyword patterns ``race*'', ``concurren*'' and ``thread*'' in the MITRE CVE database~\cite{cve-db}, we found that only 202 CVE records are relevant to concurrency-vulnerabilities out of
 the 70438 assigned CVE IDs ranging from CVE-2014-* to CVE-2018-*. In particular, we observed
that, {theoretically}, at most 4 CVE records could be detected by grey-box fuzzers 
that work on user-space programs.

As a result, {there are no practical fuzzing techniques to test \emph{ input-dependent user-space multithreaded programs} and detect bugs or vulnerabilities inside them.}
To this end, we present a dedicated grey-box fuzzing technique, 
\muzz, to reveal {bugs} by exercising \emph{input}-dependent and \emph{interleaving}-dependent paths.
We categorize the targeted \emph{multithreading-relevant bugs} into two major groups: 
\begin{enumerate}[$\bullet$,noitemsep,topsep=0pt,parsep=0pt]
    \item \textbf{concurrency-vulnerabilities} (\VulsMT): they correspond to    memory corruption vulnerabilities that occur in a multithreading context. 
    These vulnerabilities can be detected during the \emph{fuzzing} phase.
    \item \textbf{concurrency-bugs} (\BugsMT): they correspond to the bugs like data-races, atomicity-violations, deadlocks, etc. We detect them 
    by \emph{replaying} the seeds generated by \muzz with state-of-the-art concurrency-bug detectors such as \ts.
\end{enumerate}
Note that \BugsMT may not be revealed during \emph{fuzzing} since they do not necessarily
result in memory corruption crashes. In the remaining sections, when referring 
to \emph{multithreading-relevant bugs}, we always mean the combination of concurrency-bugs 
and concurrency-vulnerabilities, i.e., $\VulsMT\cup\BugsMT$.

We summarize the contributions of our work as follows:
1) We develop three novel thread-aware instrumentations for grey-box fuzzing that can distinguish the execution states caused by thread-interleavings.

\noindent 2) We optimize seed selection and execution strategies based on the runtime feedback provided by the instrumentations, which help generate more effective seeds concerning the multithreading context. 

\noindent 3) We integrate these analyses into \muzz for an effective bug hunting in
multithreaded programs. Experiments on 12 real-world programs show that \muzz 
outperforms other fuzzers like AFL and \mopt in detecting concurrency-vulnerabilities and revealing concurrency-bugs.

\noindent4) \muzz detected 8 new concurrency-vulnerabilities and 19 new concurrency-bugs, with 4 CVE IDs assigned. Considering the small portion of concurrency-vulnerabilities recorded in the CVE database, the results are promising.

\begin{algorithm}[t]
 \small
	\SetKwInOut{Input}{input}
	\SetKwInOut{Output}{output}
	\Input{program \ProgO, initial seed queue \Seeds}
	\Output{final seed queue \FinalSeeds, vulnerable seed files \CrashSeeds}
	\Prog $\leftarrow$ instrument(\ProgO) \tcp*[r]{\textcolor{blue}{instrumentation}}
	$\CrashSeeds \leftarrow~\emptyset$\; 
	\While {True} {
		t $\leftarrow$ select\_next\_seed(\FinalSeeds) \tcp*[r]{\textcolor{blue}{seed selection}}
		\mutChance $\leftarrow$ get\_mutation\_chance(\Prog, t) \tcp*[r]{\textcolor{blue}{seed scheduling}} \label{line:algo:energy}
		\For {$i\in~1\ldots \mutChance$} {
			$t'\leftarrow$ mutated\_input(t)  \tcp*[r]{\textcolor{blue}{seed mutation}}
			res $\leftarrow$ run(\Prog, {t'}, \Ncal)\tcp*[r]{\textcolor{blue}{repeated execution}}\label{line:cal_exec}
			\uIf (\tcp*[f]{\textcolor{blue}{seed triaging}}){is\_crash(\textmd{res})}{\label{line:algo:triage_start}
				$\CrashSeeds \leftarrow \CrashSeeds\cup\{{t'}\}$ \tcp*{\textcolor{teal}{report vulnerable seeds}}
			}\ElseIf {cov\_new\_trace({t', res})} {\label{line:algo:new_cov}
				$\FinalSeeds \leftarrow \FinalSeeds\oplus{t'}$ \tcp*{\textcolor{teal}{preserve ``effective'' seeds}} \label{line:algo:triage_end}
			}
		}
	}
	\caption{Grey-box Fuzzing Workflow}\label{algo:gbf}
\end{algorithm}

\section{Background and Motivation}\label{sec:motivation}

\subsection{Grey-box Fuzzing Workflow}\label{sec:gbf-workflow}

Algorithm~\ref{algo:gbf} presents the typical workflow of a grey-box fuzzer~\cite{fuzz_survey,aflfast,afl_detail}.
Given a target program \ProgO and the input seeds \Seeds, a GBF first utilizes instrumentation to track the coverage information in \ProgO. Then it enters the fuzzing loop:
1) \emph{Seed selection} decides which seed to be selected next; 2) \emph{Seed scheduling} decides how many mutations \mutChance will be applied on the selected seed $t$; 3) \emph{Seed mutation} applies mutations on seed $t$ to generate a new seed $t'$; 4) During \emph{repeated execution}, for each new seed $t'$, the fuzzer executes against it \Ncal times to get its execution statistics; 5) \emph{Seed triaging} evaluates $t'$  based on the statistics and the coverage feedback from instrumentation, to determine whether the seed leads to a {vulnerability}, or whether it is ``effective'' and should be preserved in the seed queue for subsequent fuzzing. Here, steps 3), 4), 5) are continuously processed {\mutChance} times. Notably, \Ncal times of repeated executions are necessary since a GBF needs to collect statistics such as average execution time for $t'$, which will be used to calculate mutation times \mutChance for seed scheduling in the next iteration.
In essence, the effectiveness of grey-box fuzzing relies on the feedback collected from the instrumentation. Specifically, the result of \algofont{cov\_new\_trace} (line~\ref{line:algo:new_cov}) is determined by the \emph{coverage} feedback.

\begin{figure}[t]
\begin{lstlisting}[language=c, framexleftmargin=.08\columnwidth,
xleftmargin=.02\columnwidth, xrightmargin=.01\columnwidth,
% xleftmargin=.08\columnwidth, xrightmargin=.03\columnwidth,
% linewidth=.95\columnwidth, 
linebackgroundcolor={%
\ifnum\value{lstnumber}>0\ifnum\value{lstnumber}<3\mtcolor\fi\fi
\ifnum\value{lstnumber}>9\ifnum\value{lstnumber}<14\mtcolor\fi\fi
\ifnum\value{lstnumber}>13\ifnum\value{lstnumber}<14\mtcolor\fi\fi
\ifnum\value{lstnumber}>16\ifnum\value{lstnumber}<18\mtcolor\fi\fi
}]
int g_var = -1; %\label{line:g_var_def}%
void modify(int *pv) { *pv -= 2;} %\label{line:st_mt_func_start}\hfill//\quad\circleNUMT{9}%

void check(char * buf) {
  if (is_invalid(buf)) { exit(1); } %\label{line:st_func_start}%
  else { modify((int*)buf); } %\label{line:st_func_end}%
}

char* compute(void *s_var) {
  g_var += 1; %\label{line:mt_func_start}\hfill//\quad\circleNUMT{{1}}%
  g_var *= 2; %\label{line:mt_func_assign}\hfill//\quad\circleNUMT{{2}}%
  if ((int*)s_var[0]<0) %\label{line:branch}\hfill//\quad\circleNUMT{{3}}%
     modify((int*)s_var); %\label{line:mt_func_call}\hfill//\quad\circleNUMT{4}%
  pthread_mutex_lock(&m); %\label{line:mt_func_lock}\hfill//\quad\circleNUMT{5}%
  modify(&g_var); %\label{line:locked}\hfill//\quad\circleNUMT{6}%
  pthread_mutex_unlock(&m); %\label{line:mt_func_unlock}\hfill//\quad\circleNUMT{7}%
  return (char*)s_var; %\label{line:mt_func_end}\hfill//\quad\circleNUMT{8}%
}

int main(int argc, char **argv) {
  char * buf = read_file_content(argv[1]); %\label{line:read_buf}%
  check(buf); %\label{line:call_check}%
  pthread_t T1, T2; %\label{line:main_pthread_t}%
  pthread_create(T1,NULL,compute,buf);  %\label{line:main_thread_fork_1}%
  pthread_create(T2,NULL,compute,buf+128);  %\label{line:main_thread_fork_2}%
  ......
}
\end{lstlisting}
\caption{Code segments abstracted from real-world programs. The shadow lines denote ``suspicious interleaving scope'' introduced in \S\ref{sec:preprocess}.}
\label{fig:eg1}
\end{figure}

\subsection{The Challenge in Fuzzing Multithreaded Programs and Our Solution}\label{sec:challenge}
Figure~\ref{fig:eg1} is an abstracted multithreaded program that accepts a certain input file and distributes computing jobs to threads. Practically it may behave like compressors/decompressors (e.g., \projfont{lbzip2}, \projfont{pbzip2}), image processors (e.g., \projfont{ImageMagick}, \projfont{GraphicsMagick}), encoders/decoders (e.g., \projfont{WebM}, \projfont{libvpx}), etc. After reading the input content \codefont{buf}, it does an initial validity check inside the function \codefont{check}. It exits immediately if the buffer does not satisfy certain properties. The multithreading context starts from function~\codefont{compute} (via \codefont{pthread\_create} at lines~\ref{line:main_thread_fork_1}-\ref{line:main_thread_fork_2}). It contains shared variables \codefont{s\_var} (passed from \codefont{main}) and \codefont{g\_var} (global variables), as well as the mutex primitive \codefont{m} to exclusively read/write shared variables (via \codefont{pthread\_mutex\_lock} and \codefont{pthread\_mutex\_unlock}).

With different inputs, the program may execute different segments. For example, based on the condition of statement \circleNUM{3}, which is purely dependent on the input content (i.e., different results of \texttt{buf} provided by seed files), it may or may not execute \circleNUM{4}. Therefore, {different seed files need to be generated to exercise different paths in multithreading context --- in fact, this is the starting point that we use fuzzing to generate seed files to test multithreaded programs}.

Meanwhile, in the presence of thread-interleavings, \codefont{g\_var} (initialized with -1) may also have different values.
Let us focus on different seeds' executions at two statements: \circleNUM{1}:``\texttt{g\_var+=1}'', and \circleNUM{2}: ``\codefont{g\_var*=2}''. Suppose there are two threads: T1, T2; and T1:\circleNUM{1} is executed first. Then there are at least three interleavings:
\begin{enumerate}[label={\roman*)}]
    \item\label{itm:ti1} T1:\circleNUM{1}$\rightarrow$T2:\circleNUM{1}$\rightarrow$T2:\circleNUM{2}$\rightarrow$T1:\circleNUM{2}\hfill\codefont{g\_var=4}
    \item\label{itm:ti2} T1:\circleNUM{1}$\rightarrow$T2:\circleNUM{1}$\rightarrow$T1:\circleNUM{2}$\rightarrow$T2:\circleNUM{2}\hfill\codefont{g\_var=4}
    \item\label{itm:ti3} T1:\circleNUM{1}$\rightarrow$T1:\circleNUM{2}$\rightarrow$T2:\circleNUM{1}$\rightarrow$T2:\circleNUM{2}\hfill\codefont{g\_var=2}
\end{enumerate}
After the second \circleNUM{2} is executed, the values of \codefont{g\_var} may be different (4 and 2, respectively). Worse still, since neither \circleNUMT{1} nor \circleNUMT{2} is {an \emph{atomic} operation} in the representation of the actual program binary, many more interleavings can be observed and \codefont{g\_var} will be assigned to other values.

\textbf{The challenge.} To reveal multithreading-relevant bugs, a GBF needs to generate diverse seeds that execute different paths in multithreading context (e.g., paths inside \codefont{compute}). However, \emph{existing GBFs even have difficulties in generating seeds to reach multithreading segments}.
For example, if \codefont{check} is complicated enough, most of the seeds may fail the check and exit before entering \codefont{compute} --- this is quite common due to the low quality of fuzzer-generated seeds~\cite{fuzz_survey,QSYM}. Meanwhile, \emph{even if a seed indeed executes multithreading code, it may still fail to satisfy certain preconditions to reach the problematic context.} For example, suppose \codefont{modify} contains a vulnerability that can only be triggered when \codefont{g\_var} is 2. If the fuzzer has occasionally generated a seed that executes \codefont{compute} and the condition of \circleNUM{3} is true, with no awareness of thread-interleavings, it will not distinguish different schedules between \ref{itm:ti1}, \ref{itm:ti2} and \ref{itm:ti3}. As a result, subsequent mutations on this seed will miss important feedback regarding \codefont{g\_var}, making it difficult to generate seeds that trigger the vulnerability.

To summarize, the challenge of fuzzing multithreaded programs is, existing GBFs have difficulties in generating seeds that execute multithreading context and keep thread-interleaving execution states.

\noindent \textbf{Our solution.} 
\emph{We provide fine-grained \emph{thread-aware feedback} for seed files that execute multithreading context and distinguish more such execution states}. According to \S\ref{sec:gbf-workflow}, the preservation of seeds is based on the feedback; then we can expect that the fuzzer will preserve more distinct seeds that execute multithreading code segments in the seed queue. This means that the multithreading-relevant seeds are \emph{implicitly prioritized}. Since these seeds have already passed the validity checking, the overall quality of the generated seeds is higher. The ``Matthew Effect'' helps keep the quality of seed generations for subsequent fuzzing. Essentially, this provides a biased coverage feedback on multithreading code segments (more explanations on this are available in ~\S\ref{sec:app-rationale}.

Now let us investigate what instrumentations can be improved to existing fuzzers for \emph{thread-aware feedback}.

\subsection{Thread-aware Feedback Improvements}\label{sec:improvements}

\subsubsection{Feedback to Track Thread-interleavings and Thread-context}\label{sec:improve_feedback}

The state-of-the-art GBFs, such as AFL, instrument the entry instruction of each basicblock \emph{evenly} as the basicblock's \emph{deputy}. 
We refer to this selection strategy over {deputy instructions} as \AFLIns.
\AFLIns provides coverage feedback during the dynamic fuzzing phase to explore more paths.
During repeated execution (line~\ref{line:cal_exec} in Algorithm~\ref{algo:gbf}), AFL labels a value to each \emph{transition} that {connects the deputies of two consecutively executed basicblocks}~\cite{afl_detail}. By maintaining a set of \emph{transitions} for queued seeds, \AFLIns \emph{tracks} the ``coverage'' 
of the target program. \algofont{cov\_new\_trace} (line~\ref{line:algo:new_cov} in Algorithm~\ref{algo:gbf}) checks whether a transition indicates a new path/state.

\begin{figure}[t]
	\centering
	\begin{subfloat}[]{\includegraphics[width=.30\linewidth]{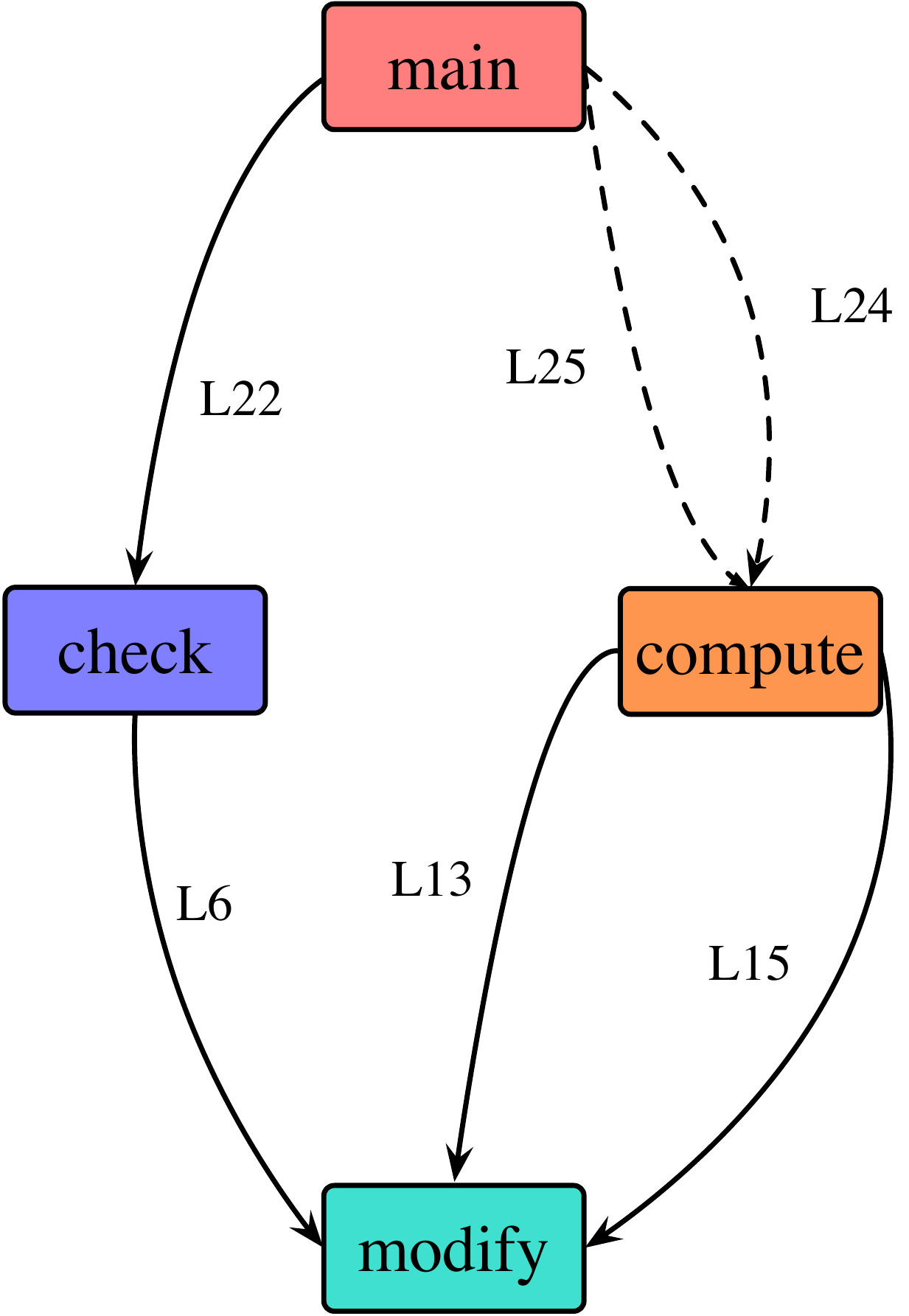}\label{subfig:callgraph}}
	\end{subfloat}
	\quad
	\begin{subfloat}[]{\includegraphics[width=.64\linewidth]{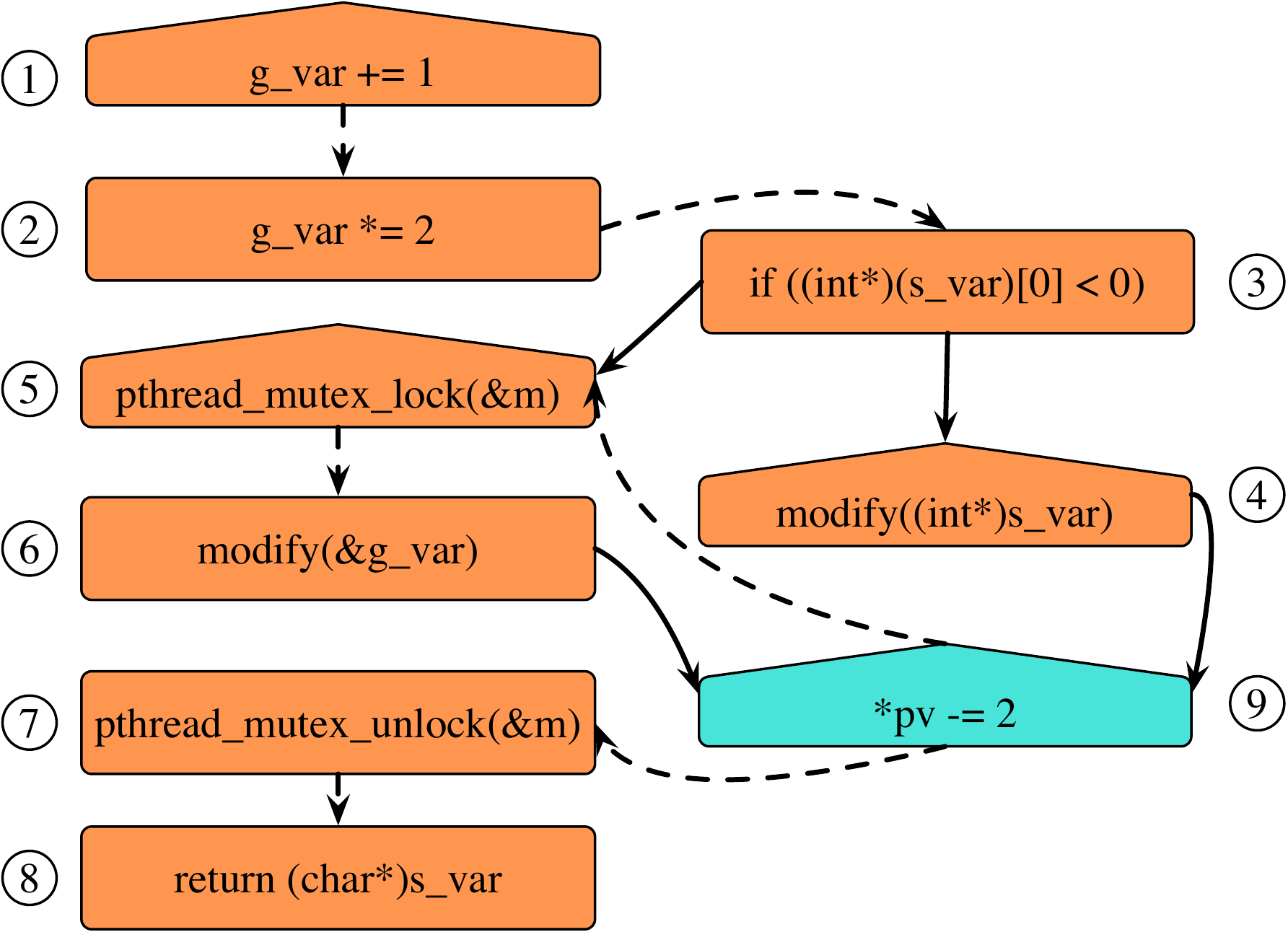}\label{subfig:transition}}
	\end{subfloat}
	\caption[]{(a) thread-aware callgraph of Figure~\ref{fig:eg1}; (b) its edge transitions across \codefont{compute} and \codefont{modify}. In (b), the arrows denote the transitions {between statements}. The pentagons denote basicblocks' entry statements; the other statements are
represented by rectangles. Their colors are consistent with function nodes in (a). Since \AFLIns only tracks branches' entry statements, only branching edges (\circleNUM{3}$\rightarrow$\circleNUM{4} and \circleNUM{3}$\rightarrow$\circleNUM{5}) and function call edges (\circleNUM{4}$\rightarrow$\circleNUM{9} and \circleNUM{6}$\rightarrow$\circleNUM{9}) are recorded --- these transitions are marked as solid arrows.}\label{fig:eg_details}
\end{figure}

Figure~\ref{subfig:transition} depicts the transitions upon executing the functions \codefont{compute} 
and \codefont{modify} in Figure~\ref{fig:eg1}. For brevity, we use \emph{source code} to illustrate the problem and use \emph{statements} to represent \emph{instructions} in assembly or LLVM IR~\cite{LLVM}.

\AFLIns works perfectly on single-threaded programs: the kept transitions can reflect both branching conditions
(e.g., \circleNUM{3}$\rightarrow$\circleNUM{4} and \circleNUM{3}$\rightarrow$\circleNUM{5})
and function calls 
(e.g., \circleNUM{4}$\rightarrow$\circleNUM{9} and \circleNUM{6}$\rightarrow$\circleNUM{9}).
However, \AFLIns cannot capture these differences among schedules \ref{itm:ti1}, \ref{itm:ti2} and \ref{itm:ti3} (c.f. \S\ref{sec:challenge}). In fact, it can only observe there is a transition \circleNUM{1}$\rightarrow$\circleNUM{1}; thus it will not prioritize this path for subsequent mutations, compared to other paths that do not even execute \codefont{compute}. The root cause of this defect lies in that AFL only tracks entry statements of basicblocks \emph{evenly}, and does not record thread identities.
Therefore, {we can add \emph{more deputy instructions} within multithreading-relevant basicblocks to provide more interleaving feedback, and add \emph{thread-context information} to distinguish different threads}.

\subsubsection{Schedule-intervention Across Executions}\label{sec:improve_diversity}
During a GBF's repeated execution procedure (line~\ref{line:cal_exec} in Algorithm~\ref{algo:gbf}), a seed may exhibit \emph{non-deterministic behaviors}: it executes different paths of the target program across executions due to randomness.
In this scenario, AFL (and other GBFs) will execute against such a seed more times than a seed with deterministic behaviors~\cite{afl_detail}. For the non-deterministic behaviors caused by scheduling-interleaving in multithreaded programs, since the execution is \emph{continuously} repeated \Ncal times, the system level environment (e.g., CPU usage, memory consumption, I/O status) is prone to be similar~\cite{posixstd,tlpi}.
This will {decrease the diversities of schedules}, and consequently reduce the overall effectiveness. For example, during a repeated execution with $\Ncal=40$, schedules \ref{itm:ti1} and \ref{itm:ti3} might occur 10 and 30 times respectively, while schedule \ref{itm:ti2} do not occur at all; in this scenario, the execution states corresponding to \ref{itm:ti2} will not be observed by the fuzzer.
Ideally, we would like the fuzzer to observe as many distinct interleavings as possible during repeated execution since that marks the potential states a seed {can exercise}. In the case of statements \circleNUM{1} and \circleNUM{2}, we hope schedules \ref{itm:ti1}, \ref{itm:ti2}, \ref{itm:ti3}
can all occur.
Therefore, {it is favorable to provide \emph{schedule interventions} to diversify the actual schedules.}
 
\begin{figure*}[ht]
    \centering
    \includegraphics[width=.85\textwidth]{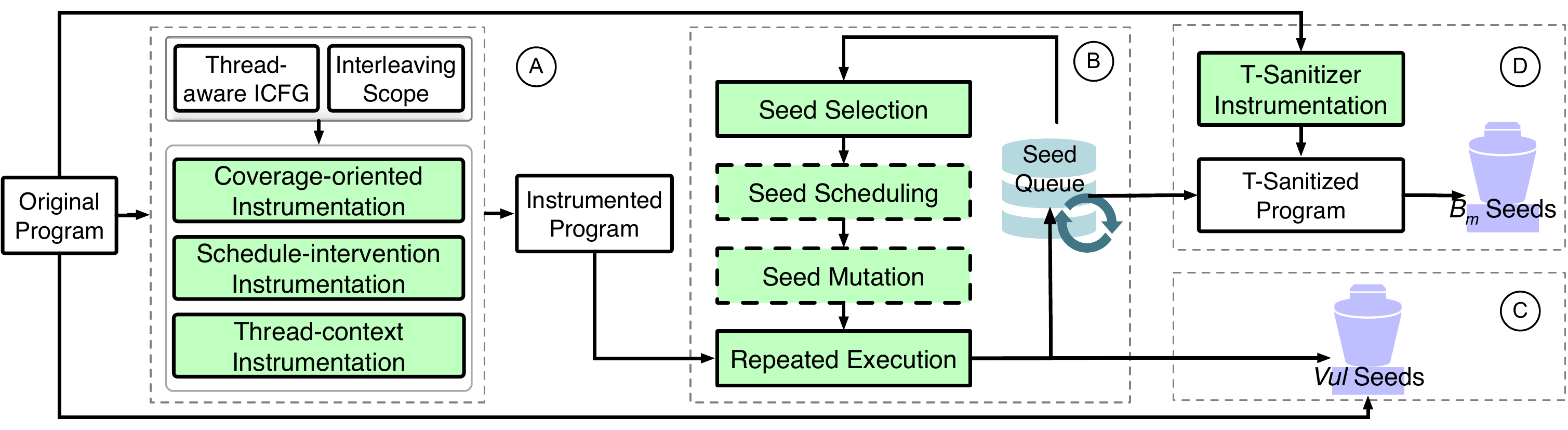}
    \caption[]{Overview of \muzz. Inputs are the original program and initial seeds (in seed queue); outputs are the seeds with vulnerabilities or concurrency-bugs. It contains four components. \PinstF (left area) does static analysis and applies thread-aware instrumentations; \PfuzzF (center area) contains the flows that proceed with dynamic fuzzing (seed scheduling and seed mutation~\cite{fuzz_survey} are the same as typical GBF flows, thus are marked dashed); \PvulF (right-bottom) denotes the vulnerability analysis applied on vulnerable seeds; and \PcbugF (right-top) is the replaying component used to reveal concurrency-bugs from the seed queue.}
    \label{fig:overview}
\end{figure*}

\section{System Overview}

Figure~\ref{fig:overview} depicts the system overview of \muzz. It contains four major components: {\PinstS} static thread-aware analysis guided instrumentations, {\PfuzzS} dynamic fuzzing, {\PvulS} vulnerability analysis, {\PcbugS} concurrency-bug revealing.

During \PinstS:\emph{instrumentation} (\S\ref{sec:instrument}), for a multithreaded program \ProgO, \muzz firstly computes thread-aware inter-procedural control flow graph (ICFG) and the code segments that are likely to interleave with others during execution~\cite{DBLP:conf/ppopp/DiS16,SuiDX16}, namely \emph{suspicious interleaving scope}, in \S\ref{sec:preprocess}. Based on these results, it performs three instrumentations inspired by \S\ref{sec:improvements}.
\begin{enumerate}[label=\arabic*)]
    \item \emph{Coverage-oriented instrumentation} (\S\ref{sec:instrument_explore}) is one kind of stratified instrumentation that assigns more deputies to suspicious interleaving scope. It is the major instrumentation
    to track thread-interleaving induced coverage.
    \item \emph{Thread-context instrumentation} (\S\ref{sec:instrument_thread_ctx}) is a type of lightweight instrumentation that distinguishes different thread identities by tracking the context of threading functions for thread-forks, locks, unlocks, joins, etc.
    \item \emph{Schedule-intervention instrumentation} (\S\ref{sec:instrument_schedule}) is a type of lightweight instrumentation at the entry of a thread-fork routine that dynamically adjusts each thread's priority. This complementary instrumentation aims to diversify interleavings by intervening in the thread schedules.
\end{enumerate}

During \PfuzzS:\emph{dynamic fuzzing} (\S\ref{sec:fuzz}), \muzz optimizes \emph{seed selection} and \emph{repeated execution} to generate more multithreading-relevant seeds. For seed selection (\S\ref{sec:seed_select}), in addition to the new coverage information provided by coverage-oriented instrumentation, \muzz also prioritizes those seeds that cover new thread-context based on the feedback provided by thread-context instrumentation. For  repeated execution (\S\ref{sec:cal_exec}), owing to the schedule-intervention instrumentation, \muzz adjusts the repeating times \Ncal, to maximize the benefit of repetitions and track the interleaved execution states.

\PvulS:\emph{Vulnerability analysis} is applied to the crashing seeds found by {dynamic fuzzing}, which reveals vulnerabilities (including {\VulsMT}). \PcbugS:\emph{concurrency-bug revealing} component reveals {\BugsMT} with the help of concurrency-bug detectors (e.g., \ts~\cite{kcc:tsan}, Helgrind~\cite{helgrind}). These two components will be explained in the evaluation section (\S\ref{sec:eval}).

\section{Static Analysis Guided Instrumentation}\label{sec:instrument}
This component includes the thread-aware static analysis and the instrumentations based on it.

\subsection{Thread-aware Static Analysis}\label{sec:preprocess}
The static analysis aims to provide lightweight thread-aware information for instrumentation and runtime feedback.

\subsubsection{Thread-aware ICFG Generation}\label{sec:tcg}
We firstly apply an inclusion-based pointer analysis~\cite{Andersen94programanalysis} on the target program. The points-to results are used to resolve the def-use flow of thread-sharing variables and indirect calls to reconstruct the ICFG. By taking into account the semantics of threading APIs (e.g., POSIX standard Pthread, the OpenMP library), we get an ICFG that is aware of the following multithreading information:
\begin{enumerate}[label=\arabic*)]
    \item \TStart is the set of program sites that call thread-fork functions. This includes the explicit call to \algofont{pthread\_create}, the \algofont{std::thread} constructor that internally uses \algofont{pthread\_create}, or the ``parallel pragma'' in OpenMP. The \emph{called} functions, denoted as \FSStart, are extracted from the semantics of these forking sites.
    \item \TEnd contains call sites for functions that mark the end of a multithreading context. It includes the call sites of the pthread APIs such as \algofont{pthread\_join}, 
    \algofont{pthread\_exit}, etc.
    \item \TLock is the set of sites that call thread-lock functions such as \algofont{pthread\_mutex\_lock}, \algofont{omp\_set\_lock}, etc.
    \item \TUnLock is the set of sites that call thread-unlock functions like \algofont{pthread\_mutex\_unlock}, \algofont{omp\_unset\_lock}, etc.
    \item \TSharedVar is the set of variables shared among different threads. This includes global variables and those variables that are passed from thread-fork sites (e.g., \TStart).
\end{enumerate}

\subsubsection{Suspicious Interleaving Scope Extraction}\label{sec:extract}
Given a program that may run simultaneously with multiple threads, we hope the instrumentation to collect execution states to reflect the interleavings. However, instrumentation introduces considerable overhead to the original program, especially when it is applied intensively throughout the whole program.
Fortunately, with the static information provided by the thread-aware ICFG, we know that thread-interleavings may only happen on {some specific program statements}; therefore, the instrumentation can stress these statements. We hereby use \mtiscope to denote the set of these statements and term it as \emph{suspicious interleaving scope}. 
\mtiscope is determined according to the following three conditions.
\begin{enumerate}[{\bf C1}]
    \item\label{cond:forkjoin} The statements should be executed after one of \TStart, while \TEnd is not encountered yet.
    \item\label{cond:lock} The statements can only be executed before the invocation of \TLock and after the invocation of \TUnLock.
    \item\label{cond:sharedvar} The statements should read or write at least one of the shared variables by different threads.
\end{enumerate}

\ref{cond:forkjoin} excludes the statements irrelevant to multithreading. These statements can be prologue code that does the validity check (e.g., \codefont{check} in Figure~\ref{fig:eg1}), or the epilogue that post-processes the inputs or deals with error handlings.
\ref{cond:lock} prevents the statements that are protected by certain locks from being put into \mtiscope.
\ref{cond:sharedvar} is necessary since the interleavings will not affect the shared states if the segment involves no shared variables. This condition is determined by observing whether the investigated statement contains a variable data dependent on \TSharedVar (based on pointer analysis). We provide a separate preprocessing procedure to exclude cases where there are only \emph{read} operations on shared variables.

Note that \mtiscope is used to emphasize multithreading-relevant paths via instrumentations for state exploration during fuzzing. Therefore the conditions are different from the constraints required by static models (e.g.,  may-happen-in-parallel~\cite{DBLP:conf/ppopp/DiS16,SuiDX16}) or dynamic concurrency-bug detection algorithms (e.g., happens-before~\cite{pldi09_fasttrack} or lockset~\cite{lockset_SavageABNS97}).

In Figure~\ref{fig:eg1}, according to the call \codefont{pthread\_create} at Lines~\ref{line:main_thread_fork_1} and \ref{line:main_thread_fork_2}, $\FSStart=\{\codefont{compute}\}$. \muzz then gets all the functions that may be called by functions inside \FSStart, i.e., \{\codefont{modify},\codefont{compute}\} and according to \ref{cond:forkjoin} the scope \mtiscope comes from Lines~\ref{line:g_var_def}, \ref{line:st_mt_func_start},
\ref{line:mt_func_start}$-$\ref{line:mt_func_end}. 
Inside these functions, we check the statements that are outside \codefont{pthread\_mutex\_lock} and \codefont{pthread\_mutex\_unlock} based on \ref{cond:lock}: Line~\ref{line:locked} should be excluded from \mtiscope. 
According to \ref{cond:sharedvar}, we exclude the statements that do not access or modify the shared variables \codefont{g\_var}, \codefont{s\_var},
which means Lines~\ref{line:mt_func_lock} and \ref{line:mt_func_unlock} should also be excluded. In the end, the scope is determined as
$\mtiscope=\{\ref{line:g_var_def},\ref{line:st_mt_func_start},\ref{line:mt_func_start},\ref{line:mt_func_assign},\ref{line:branch}, \ref{line:mt_func_call},\ref{line:mt_func_end}\}$.
Note that although \codefont{modify} can be called in a single-threading site inside \codefont{check} (Line~\ref{line:st_func_end}), we still conservatively include it in \mtiscope. The reason is that it \emph{might be} called within multithreading contexts (Line~\ref{line:mt_func_call} and Line~\ref{line:locked})  --- \codefont{modify} is protected by mutex \codefont{m} at Line~\ref{line:locked} while unprotected at Line~\ref{line:mt_func_call}.
It is worth noting that line~\ref{line:locked}, although protected by \codefont{m}, may still happen-in-parallel~\cite{DBLP:conf/ppopp/DiS16,SuiDX16} with lines~\ref{line:mt_func_start} and ~\ref{line:mt_func_assign}. However, since lines~\ref{line:mt_func_start} and~\ref{line:mt_func_assign} have already been put in \mtiscope, we consider it sufficient to help provide more feedback to track thread-interleavings, with line~\ref{line:locked} excluded from \mtiscope. 
Overall, the static analysis is lightweight. For example, the pointer analysis is flow- and context-insensitive
; extraction of thread-aware results such as \FSStart (in \ref{cond:forkjoin}) and \TSharedVar (in \ref{cond:sharedvar}) are over-approximated in that the statically calculated sets may be larger than the actual sets; \ref{cond:lock} may aggressively exclude several statements that involve interleavings. The benefit, however, is that it makes our analysis scalable to large-scale real-world programs.

\subsection{Coverage-oriented Instrumentation}\label{sec:instrument_explore}
With the knowledge of \mtiscope, we can instrument more deputy \emph{instructions} (corresponding to statements in source code) inside the scope than the others, for exploring new transitions. However, it is still {costly} to instrument on \emph{each instruction} inside \mtiscope since this may significantly reduce the overall execution speed of the target programs. It is also \emph{unnecessary} to do so --- although theoretically, interleavings may happen everywhere inside \mtiscope, many interleavings are not important because they do not change the values of shared variables in practice. This means that we can {skip} some instructions for instrumentation, or equivalently instrument them \emph{with a probability}.
We still instrument, despite less, on segments outside \mtiscope for \emph{exploration} purposes~\cite{fuzz_survey}. For example, in Figure~\ref{fig:eg1}, we apply instrumentation on \codefont{check}, just in case the initial seeds are all rejected by the validity check and no intermediate feedback are available at all, making the executions extremely difficult to even enter \codefont{compute}. Similarly, we can also {selectively} instrument some instructions outside \mtiscope.

\subsubsection{Instrumentation Probability Calculation}
The goal of calculating instrumentation probabilities is to strike a balance between execution overhead and feedback effectiveness by investigating code segments' complexity of the target programs.
First of all, \muzz calculates a \emph{base instrumentation probability} according to \emph{cyclomatic complexity}~\cite{macabecc}, based on the fact that bugs or vulnerabilities usually come from functions with higher cyclomatic complexity~\cite{vul_metric,fuzz_vul_metric}. For each function $\aFUNC$, we calculate the complexity value: $M_c(\aFUNC) = E(\aFUNC) - N(\aFUNC) + 2$ where $N(f)$ is the number of nodes (basicblocks) and $E(\aFUNC)$ is the number of edges in the function's control flow graph. Intuitively, this value determines the complexity of the function across its basicblocks. As 10 is considered to be the preferred upper bound of $M_c$~\cite{macabecc}, we determine the base probability as:
\begin{equation}
    \ccr{\aFUNC} = \min\big\{\frac{E(\aFUNC)-N(\aFUNC)+2}{10}, ~1.0\big\}
\end{equation}
We use $\stpn$ as the probability to {selectively} instrument on the entry instruction of a basicblock that is \emph{entirely outside} suspicious interleaving scope, i.e., {none of} the instructions inside the basicblock belong to \mtiscope. Here, $\stpn$ is calculated as:
\begin{equation}
    \stp{\aFUNC} = \min\big\{\ccr{\aFUNC}, ~\stpnO\big\}
\end{equation}
where $0<\stpnO<1$. Empirically, \muzz sets $\stpnO=0.5$.

Further, for each basicblock \aBB inside the given function~\aFUNC, we calculate the total number of instructions $N(\aBB)$, and the total number of memory operation instructions $N_{\mathbf{m}}(\aBB)$ (e.g., load/store, \algofont{memcpy}, \algofont{free}). 
Then for the instructions within~\mtiscope, the instrumentation probability is calculated as:
\begin{equation}
    \mtp{\aFUNC}{\aBB} = \min\big\{\ccr{\aFUNC}\cdot\frac{N_{\mathbf{m}}(\aBB)}{N(\aBB)}, ~\mtpnO \big\}
\end{equation}
where $\mtpnO$ is a factor satisfying $0<\mtpnO<1$ and defaults to $0.33$. The rationale of $\frac{N_{\mathbf{m}}(\aBB)}{N(\aBB)}$ is that vulnerabilities usually result from memory operation instructions~\cite{fuzz_survey}, and executions on more such operations deserve more attention.

\subsubsection{Instrumentation Algorithm}
The coverage-oriented instrumentation algorithm is described in Algorithm~\ref{algo:inst_explore}. It traverses functions in the target program $P$. For each basicblock \aBB in function \aFUNC, \muzz firstly gets the intersection of the instructions inside both \aBB and \mtiscope. If this intersection $\mtiscope\!(\aBB)$ is empty, it instruments the {entry instruction} of \aBB with a probability of \emph{\stpn(\aFUNC)}. Otherwise, 1) for the entry instruction in \aBB, \muzz always instruments it (i.e., with probability 1.0); 2) for the other instructions, if they are inside \mtiscope, \muzz instruments them with a probability of \mtp{\aFUNC}{\aBB}. We will refer to our selection strategy over deputy instructions as \MTIns. As a comparison, \AFLIns always instruments evenly at the entry instructions of all the basicblocks.

\begin{algorithm}[t]
\small
  \SetKwInOut{Input}{input}
  \SetKwInOut{Output}{output}
\Input{target program \algofont{P}, and suspicious interleaving scope~\mtiscope}
\Output{program \algofont{P} instrumented with \MTIns deputies}
\For{{\aFUNC} $\in$ \algofont{P}}{
  \For{{\aBB} $\in$ \aFUNC}{
    $\mtiscope(\aBB) = \mtiscope\cap~\aBB$\;
    \uIf{$\mtiscope(\aBB)~!\!=\emptyset$}{
      \For{\aINSTR $\in$ \aBB}{
        \uIf{{is\_entry\_instr(\aINSTR, \aBB)}}{
          P $\leftarrow$ instrument\_cov(P, \aINSTR, 1.0)\;
        }
        \ElseIf{$\aINSTR\in\mtiscope$}{
          P $\leftarrow$ instrument\_cov\big(P, \aINSTR, \mtp{\aFUNC}{\aBB}\big)\;
        }
      }      
    }\Else{
      \For{\aBB $\in$ \aFUNC}{
        \aINSTR = get\_entry\_instr(\aBB)\;
        P $\leftarrow$ instrument\_cov\big(P, \aINSTR, \stp{\aFUNC}\big)\;
      }
    }
  }
}
\caption{Coverage-oriented Instrumentation}\label{algo:inst_explore}
\end{algorithm}

For the example in Figure~\ref{fig:eg1}, since the lines~\ref{line:read_buf}-\ref{line:main_thread_fork_2} and line~\ref{line:st_func_start} are out of \mtiscope, we can expect \MTIns to instrument fewer entry statements on their corresponding basicblocks. Meanwhile, for the statements inside \mtiscope, \MTIns may instrument other statements besides the entry statements. For example, ~\circleNUM{1} is the entry statement thus it must be instrumented; statement~\circleNUM{2} may also be instrumented (with a probability) --- if so, transition~\circleNUM{1}$\rightarrow$\circleNUM{2} can be tracked.

\subsection{Threading-context Instrumentation}\label{sec:instrument_thread_ctx}
We apply threading-context instrumentation to distinguish thread identities for additional feedback. This complements \emph{coverage-oriented instrumentation} since the latter is unaware of thread IDs. The context is collected at the call sites of $\FSThread=\{\TLock, \TUnLock, \TEnd\}$, each of which has the form $\threadCtxn=\threadCtx{\iloc}{\ntid}$, where \iloc is the labeling value of deputy instruction executed before this call site
, and \ntid is obtained by getting the value of the key identified by current thread ID from the ``thread ID map'' collected by the instrumented function \rtifunc (to be explained in \S\ref{sec:instrument_schedule}).
Given an item $F$ in \FSThread, we keep a sequence of context $\langle\threadCtxn{_1}(F),\ldots,\threadCtxn{_n}(F)\rangle$,$F\in\FSThread$. At the end of each execution, we calculate a hash value $H(F)$ for item $F$. The tuple $\tctxSign=\big\langle H(\TLock),H(\TUnLock),H(\TEnd)\big\rangle$ is a \emph{context-signature} that determines the overall thread-context of a specific execution. Essentially, this is a \emph{sampling} on threading-relevant APIs
to track the thread-context of a specific execution. As we shall see in \S\ref{sec:seed_select}, the occurrence of \tctxSign determines the results of \algofont{cov\_new\_mt\_ctx} during seed selection.

In Figure~\ref{fig:eg1}, each time when \codefont{pthread\_mutex\_lock}$\in\TLock$ is called, \muzz collects the deputy instruction prior to the corresponding call site (e.g., \circleNUM{3}) and the thread ID label (e.g., T1) to form the tuple (e.g., $\langle \circleNUM{3},T1\rangle$); these tuples form a sequence for {\TLock}, and a hash value $H(\TLock)$ will be calculated eventually. Similar calculations are applied for \codefont{pthread\_mutex\_unlock} and \codefont{pthread\_join}.

\subsection{Schedule-intervention Instrumentation}\label{sec:instrument_schedule}
When a user-space program does not specify any scheduling policy or priority, the operating system determines the actual schedule dynamically~\cite{tlpi,posixstd}. Schedule-intervention instrumentation aims to diversify the thread-interleavings  to collaborate with coverage-oriented and thread-context instrumentations. 
This instrumentation should be general enough to work for different multithreaded programs and extremely lightweight to keep runtime overhead minimal.

\begin{algorithm}[t]
\small
  \SetKwInOut{Input}{input}
  \SetKwInOut{Output}{output}
  \Input{seed queue~\Seeds, seed $t$ at queue front}
  \Output{whether $t$ will be selected in this round}
  \uIf{has\_new\_mt\_ctx(\Seeds) \textmd{or} has\_new\_trace(\Seeds)}{
      \uIf{cov\_new\_mt\_ctx($t$)}{
        \Return{true}\;
      } \uElseIf{{cov\_new\_trace($t$)}}{
        \Return{{select\_with\_prob(\probYNT)}}\;
      } \Else{
        \Return{{select\_with\_prob(\probYNN)}}\;
      }
  }\Else{
    \Return{{select\_with\_prob(\probNNN)}}\;
  }
  \caption{\algofont{select\_next\_seed} Strategy}\label{algo:select_seed}
\end{algorithm}

POSIX compliant systems such as Linux, FreeBSD
usually provide APIs to control the low-level process or thread schedules~\cite{posixstd,tlpi}.
In order to intervene in the interleavings during the execution of the multithreading segments, we resort to the POSIX API \algofont{pthread\_setschedparam} to adjust the thread priorities with an instrumented function named \rtifunc that will be invoked during fuzzing.
This function does two tasks:
\begin{enumerate}[a)]
    \item During repeated execution (\S\ref{sec:cal_exec}), whenever the thread calls \rtifunc, it updates the scheduling policy to \emph{SCHED\_RR}, and assigns a {ranged random value} to its priority.
    This value is \emph{uniformly distributed random} and diversifies the actual schedules across different threads. With this intervention, we try to approximate the goal in \S\ref{sec:improve_diversity}.
    \item For each newly mutated seed file, it calls \algofont{pthread\_self} in the entry of \FSStart to collect the thread IDs. It has two purposes: 1) it informs the fuzzer that the current seed is multithreading-relevant; 2) based on the invocation order of \rtifunc, each thread can be associated with a unique ID \ntid starting from $1,2,\ldots$, which composes ``thread ID map'' and calculates thread-context in \S\ref{sec:instrument_thread_ctx}.
\end{enumerate}

\section{Dynamic Fuzzing}\label{sec:fuzz}
The dynamic fuzzing loop follows the workflow of a typical GBF described in Algorithm~\ref{algo:gbf}. To improve the feedback on multithreading context, we optimize seed selection (\S\ref{sec:seed_select}) and repeated execution (\S\ref{sec:cal_exec}) for fuzzing multithreaded programs, based on the aforementioned instrumentations.

\subsection{Seed Selection}\label{sec:seed_select}

Seed selection decides which seeds to be mutated next. In practice, this problem is reduced to: {when traversing seed queue $\Seeds$, whether the seed \emph{$t$} at the queue front will be selected for mutation}. Algorithm~\ref{algo:select_seed} depicts our solution. The intuition is that we prioritize those seeds \emph{with new (normal) coverage} or \emph{covering new thread-context}.

In addition to following AFL's strategy by using \algofont{has\_new\_trace(\Seeds)} to check whether there exists a seed, $s$, in \Seeds that covers a new transition (i.e., \algofont{cov\_new\_trace(s)}==true), \muzz also uses \algofont{has\_new\_mt\_ctx(\Seeds)} to check whether there exists a seed in \Seeds with a new thread-context (\tctxSign). If either is satisfied, it means there exist some ``interesting seeds'' in the queue.
Specifically, if the current seed $t$ covers a new thread-context, the algorithm directly returns true. If it covers a new trace, it has a probability of \probYNT to be selected; otherwise, the probability is \probYNN. On the contrary, if no seeds in \Seeds are interesting, the algorithm selects $t$ with a probability of \probNNN. Analogous to AFL's seed selection strategy~\cite{afl_detail}, \muzz sets $\probYNT=0.95$, $\probYNN=0.01$, $\probNNN=0.15$.

As to the implementation of $\algofont{cov\_new\_mt\_ctx(t)}$, we track the thread-context of calling a multithreading API in $\FSThread=\{\TEnd, \TLock, \TUnLock\}$ (c.f. \S\ref{sec:instrument_thread_ctx}) and check whether the  context-signature \tctxSign has been met before ---  when \tctxSign is new, $\algofont{cov\_new\_mt\_ctx(t)}$=true; otherwise, $\algofont{cov\_new\_mt\_ctx(t)}$=false. Note that $\algofont{cov\_new\_trace(t)}$==true does not imply $\algofont{cov\_new\_mt\_ctx(t)==true}$. The reason is that (1) we cannot instrument inside the body of threading API functions (as they are ``external functions'') inside \FSThread, hence \algofont{cov\_new\_trace} cannot track the transitions; (2) $\algofont{cov\_new\_mt\_ctx}$ also facilitates the thread IDs that $\algofont{cov\_new\_trace}$ is unaware of.

\subsection{Repeated Execution}\label{sec:cal_exec}
Multithreaded programs introduce non-deterministic behaviors when different interleavings are involved.
As mentioned in \S\ref{sec:improve_diversity}, for a seed with non-deterministic behaviors, a GBF typically repeats the execution on the target program against the seed for more times. 
With the help of~\rtifunc (c.f. \S\ref{sec:instrument_schedule}), we are able to tell {whether or not the exhibited non-deterministic behaviors result from thread-interleavings}. In fact, since we focus on multithreading only, based on the thread-fork information kept by \rtifunc, the fuzzer can distinguish the seeds with non-deterministic behaviors purely by checking whether the executions exercise multithreading context. Further, if previous executions on a seed induce more distinct values of \tctxSign (the number of these values for a provided seed $t$ is denoted as \Nmt(t)), we know that there must exist more thread-interleavings. To determine the repeating times~\Ncal applied on $t$, we rely on $\Nmt(t)$.
In AFL, the repeating times on $t$ is:
\begin{equation}
    \Ncal(t) = \NcalO + \NcalV\cdot\NcalB,\quad\NcalB\in\{0,1\} \label{eq:afl_cal}
\end{equation}
where \NcalO is the initial repeating times, \NcalV is a constant as the ``bonus'' times for non-deterministic runs. \NcalB=0 if none of the \NcalO executions exhibit non-deterministic behaviors; otherwise \NcalB=1. We augment this to fit for multithreading setting. 
\begin{equation}
    \Ncal(t) = \NcalO + \texttt{min}\big\{\NcalV, \NcalO\cdot\Nmt(t)\big\} \label{eq:mtfuzz_cal}
\end{equation}
In both AFL and \muzz, $\NcalO=8$, $\NcalV=32$.
For all the \Ncal executions, we track their execution traces and count how many different states it exhibits. The rationale of adjusting \Ncal is that, in real-world programs {the possibilities of thread-interleavings can vary greatly for different seeds}. For example, a seed may exhibit non-deterministic behaviors when executing~\codefont{compute} in Figure~\ref{fig:eg1} (e.g., races in~\codefont{g\_var}), but it exits soon after failing an extra check inside \codefont{compute} (typically, exit code >0). For sure, it will exhibit fewer non-deterministic behaviors than a seed that is concurrently processed and the program exits normally (typically, exit code =0). 

\subsection{Complementary Explanations}
\label{sec:app-rationale}

Here we provide some explanations to show why \muzz's static and dynamic thread-aware strategies help to improve the overall fuzzing effectiveness.

\textbf{1) Mutations on multithreading-relevant seeds are more valuable for seed mutation/generation.}
Multithreading-relevant seeds themselves have already passed validity checks of the target program. Compared to a seed that cannot even enter the thread-fork routines, it is usually much easier to generate a multithreading-relevant seed mutant from an existing multithreading-relevant seed. This is because the mutation operations (e.g., bitwise/bytewise flips, arithmetic adds/subs) in grey-box fuzzers are rather random and it is rather difficult to turn an invalid seed to be valid. Therefore, from the \emph{mutation's perspective}, we prefer multithreading-relevant seeds to be mutated.

\textbf{2) \muzz can distinguish more multithreading-relevant states.}
For example, in Figure~\ref{fig:eg1}, it can distinguish transitions \circleNUM{1}$\rightarrow$\circleNUM{1}$\rightarrow$\circleNUM{2}$\rightarrow$\circleNUM{2} and \circleNUM{1}$\rightarrow$\circleNUM{2}$\rightarrow$\circleNUM{1}$\rightarrow$\circleNUM{2}. Then when \emph{two different seeds} exercise the two transitions, \muzz is able to preserve both seeds. However, other GBFs such as AFL cannot observe the difference.
Conversely, when we provide less feedback for seeds that do not involve multithreading, \muzz can distinguish less of these states and put less multithreading-\emph{irrelevant} seeds in the seed queue.

\textbf{3) Large portions of multithreading-relevant seeds in the seed queue benefit subsequent mutations.}
Suppose at some time of fuzzing, both \muzz and AFL preserve 10 seeds (\testsALL=10), and \muzz keeps 8 multithreading-relevant seeds (\testsMT=8) while AFL keeps 6 (\testsMT=6). Obviously, the probability of picking \muzz generated multithreading-relevant seeds (80\%) is higher than AFL's (60\%). After this iteration of mutation, more seed mutants in \muzz are likely multithreading-relevant. The differences of seed quality (w.r.t. relevance to multithreading) in the seed queue can be amplified with more mutation iterations. For example, finally \muzz may keep 18 multithreading-relevant seeds (\testsMT=18), and 10 other seeds, making \testsALL=28 and \testsRatio=$\frac{18}{28}=64.3\%$; while AFL keeps 12 multithreading-relevant seeds (\testsMT=12) and 14 other seeds, making \testsALL=26 and \testsRatio=$\frac{12}{26}=46.2\%$.

Properties \textbf{1)}, \textbf{2)} and \textbf{3)} collaboratively affects the fuzzing effectiveness in a ``closed- loop''. Eventually, both \testsMT and \testsRatio in \muzz are likely to be bigger than those in AFL. Owing to more multithreading-relevant seeds in the queue and property \textbf{1)}, we can expect that:
\begin{enumerate}[a)]
    \item concurrency-vulnerabilities are more likely to be detected with the new proof-of-crash files mutated from multithreading-relevant files from the seed queue.
    \item concurrency-bugs are more likely to be revealed with the (seemingly normal) files in the seed queue that violate certain concurrency conditions.
\end{enumerate}

Providing more feedback for multithreading-relevant segments essentially provides a biased coverage criterion to specialize fuzzing on multithreaded programs. Other specialization techniques, such as the \emph{context-sensitive instrumentation} used by Angora~\cite{Angora}, or the \emph{typestate-guided instrumentation} in UAFL~\cite{Wang2020Typestate}, provide similar solutions and achieve inspiring results. The novelty of \muzz lies in that we facilitate the multithreading-specific features as feedback to innovatively improve the seed generation quality. It is worth noting that our solution only needs lightweight thread-aware analyses rather than deep knowledge of multithreading/concurrency; thus, it can scale to real-world software.  

\begin{table*}[ht]
\caption{Static statistics of the 12 evaluated benchmarks; meanings of the columns are explained in \S\ref{sec:bm_stats}.}
\label{tbl:bm_stats}
\renewcommand\arraystretch{0.9}
\footnotesize 
\centering
\begin{tabular}{c ccrrrrrr}
\toprule
\textbf{ID} &\textbf{Project}  & \textbf{Command Line Options}  & \textbf{\begin{tabular}[c]{@{}c@{}}Binary\\ Size\end{tabular}} &  \textbf{$T_{pp}$} & \textbf{$N_{b}$} & \textbf{$N_i$} & \textbf{$N_{ii}$} & \textsf{$\frac{N_{ii}-N_b}{N_b}$} \\
\midrule
\textbf{lbzip2-c} & lbzip2-2.5  & \pbin{lbzip2} -k -t -9 -z -f -n4 FILE & 377K & 7.1s & 4010 & 24085 & 6208 & 54.8\% \\  
\textbf{pbzip2-c} & pbzip2-v1.1.13  & \pbin{pbzip2} -f -k -p4 -S16 -z FILE & 312K & 0.9s & 2030 & 8345 & 2151 & 6.0\% \\  
\textbf{pbzip2-d} & pbzip2-v1.1.13  & \pbin{pbzip2} -f -k -p4 -S16 -d FILE & 312K & 0.9s & 2030 & 8345 & 2151 & 6.0\% \\ 
\textbf{pigz-c} & pigz-2.4  & \pbin{pigz} -p 4 -c -b 32 FILE & 117K & 5.0s & 3614  & 21022 & 5418 & 49.9\% \\ 
\textbf{pxz-c} & pxz-4.999.9beta  & \pbin{pxz} -c -k -T 4 -q -f -9 FILE & 42K & 1.2s & 3907 & 30205 & 7877 & 101.6\% \\ 
\textbf{xz-c} & XZ-5.3.1alpha  & \pbin{xz} -9 -k -T 4 -f FILE & 182K & 8.4s & 4892 & 34716 & 8948 & 82.9\% \\ 
\textbf{gm-cnvt} &  GraphicsMagick-1.4  & \pbin{gm} convert -limit threads 4  FILE out.bmp & 7.6M & 224.4s & 63539 & 383582 & 98580 & 55.1\% \\ 
\textbf{im-cnvt} & ImageMagick-7.0.8-7  & \pbin{convert} -limit thread 4 FILE out.bmp & 19.4M & 434.2s & 128359 & 778631 & 200108 & 55.9\% \\ 
\textbf{cwebp} & libwebp-1.0.2  & \pbin{cwebp} -mt FILE -o out.webp & 1.8M & 56.3s & 12117 & 134824 & 33112 & 173.3\%\\ 
\textbf{vpxdec} & libvpx-v1.3.0-5589  & \pbin{vpxdec} -t 4 -o out.y4m FILE & 3.8M & 431.6s & 31638 & 368879 & 93400 & 195.2\% \\ 
    \textbf{x264} & x264-0.157.2966 & \pbin{x264} --threads=4 -o out.264 FILE & 6.4M & 1701.0s & 38912 & 410453 & 103926 & 167.1\% \\ 
\textbf{x265} & x265-3.0\_Au+3 & \pbin{x265} --input FILE --pools 4 -F 2 -o & 9.7M & 78.3s & 22992 & 412555 & 89408 & 288.9\% \\ 
\bottomrule
\end{tabular}
\end{table*}

\section{Evaluation}\label{sec:eval}
We implemented \muzz upon SVF~\cite{Sui:2016:SVF}, AFL~\cite{afl_detail} , and ClusterFuzz~\cite{clusterfuzz}. 
The thread-aware ICFG construction leverages SVF's inter-procedural value-flow 
analysis. The instrumentation and dynamic fuzzing strategies lay inside AFL's LLVM module. The vulnerability analysis and concurrency-bug replaying components rely on ClusterFuzz's crash analysis module.
We archive our supporting materials 
at \url{https://sites.google.com/view/mtfuzz}. The archive includes initial seeds for fuzzing, the detected concurrency-vulnerabilities and concurrency-bugs, the implementation details, and other findings during evaluation.

Our evaluation targets the following questions:
\begin{enumerate}[label={\bf RQ\arabic*}]
    \item Can \muzz generate more effective seeds that execute multithreading-relevant program states?
    \item What is the capability of \muzz in detecting concurrency-vulnerabilities (\VulsMT)?
    \item What is the effect of using \muzz generated seeds to reveal concurrency-bugs (\BugsMT) with bug detectors?
\end{enumerate}

\subsection{Evaluation Setup}

\subsubsection{Settings of the grey-box fuzzers}
We use the following fuzzers during evaluation.
\begin{enumerate}[1)]
\item \textbf{\muzz} is our full-fledged fuzzer that applies all the thread-aware strategies in \S\ref{sec:instrument} and \S\ref{sec:fuzz}.
\item \textbf{\mafl} is a variant of \textbf{\muzz}. It differs from \muzz only on the coverage-oriented instrumentation --- \mafl uses \AFLIns while \muzz uses \MTIns. We compare \mafl with \muzz to demonstrate the effectiveness of \MTIns, and compare \mafl with AFL to stress other strategies.
\item \textbf{AFL} is by far the most widely-used GBF that facilitates general-purpose \AFLIns instrumentation and fuzzing strategies. It serves as the baseline fuzzer.
\item \textbf{\mopt}~\cite{mopt} is the recently proposed general-purpose fuzzer that leverages adaptive mutations to increase the overall fuzzing efficiency. It is claimed to be able to detect 170\% more vulnerabilities than AFL in fuzzing (single-thread) programs.
\end{enumerate}

\subsubsection{Statistics of the evaluation dataset}\label{sec:bm_stats}
The dataset for evaluation consists of the following projects.
\begin{enumerate}[1),noitemsep,leftmargin=*,topsep=0pt,parsep=0pt,labelindent=0pt]
    \item Parallel compression/decompression utilities including \projfont{pigz}, \projfont{lbzip2}, 
			\projfont{pbzip2}, \projfont{xz} and \projfont{pxz}. These tools have been present in 
			GNU/Linux distributions for many years and are integrated
into the GNU \textsf{tar} utility.
    \item \projfont{ImageMagick} and \projfont{GraphicsMagick} are two widely-used software suites to display, convert, and edit image files.
    \item \projfont{libvpx} and \projfont{libwebp} are two WebM projects for VP8/VP9 and WebP codecs. They are used by popular browsers like Chrome, Firefox, and Opera.
    \item \projfont{x264} and \projfont{x265} are the two most established video encoders for
H.264/AVC and HEVC/H.265, respectively.
\end{enumerate}

All these projects' single-thread functionalities have been intensively tested by mainstream GBFs such as AFL. We try to use their {latest} versions at the time of our evaluation; the only exception is \projfont{libvpx}, which we use version v1.3.0-5589 to reproduce the ground-truth vulnerabilities and concurrency-bugs. Among the 12 multithreaded programs, \projfont{pxz}, \projfont{GraphicsMagick}, and \projfont{ImageMagick} use OpenMP library, while the others use native PThread.

Table~\ref{tbl:bm_stats} lists the statistics of the benchmarks. The first two 
columns show the benchmark IDs and their host projects. The next column specifies the command-line options. In particular, four working threads are specified to enforce the program to run in multithreading mode.

The rest of the columns are the static statistics.  Column ``Binary Size'' 
calculates the sizes of the instrumented binaries. Column $T_{pp}$ records the 
 preprocessing time of static analysis (c.f.~\S\ref{sec:preprocess}). Among the 12 benchmarks, \emph{vpxdec} takes the longest time of approximately 30 minutes. Columns $N_b$, $N_i$, and $N_{ii}$ 
depict the number of basicblocks, the number of total instructions, and the number 
of deputy instructions for \MTIns (c.f.~\S\ref{sec:instrument_explore}), 
respectively. Recall that \AFLIns instruments evenly over entry instructions 
of all basicblocks, hence $N_b$ also denotes the number of deputy instructions in AFL, \mafl, and \mopt. 
The last column, $\frac{N_{ii}-N_b}{N_b}$, is the ratio of more instructions \muzz instrumented versus AFL (or \mafl, \mopt). This ratio ranges from 6.0\% (\projfont{pbzip2-c} or \projfont{pbzip2-d}) to 288.9\% (\projfont{x265}). Fortunately, in practice, this does not proportionally increase the runtime overhead. Many aspects can affect this metric, including the characteristics of the target programs, the precision of the applied static analysis, and the empirically  specified thresholds \stpnO and \mtpnO.

\noindent \textbf{Fuzzing Configuration}
The experiments are conducted on four Intel(R) Xeon(R) Platinum
8151 CPU@3.40GHz workstations with 28 cores, each of which runs a 64-bit Ubuntu 18.04 LTS; the evaluation upon a specific benchmark is conducted on one machine.
To make fair comparisons, \muzz, \mafl and AFL are executed in their ``fidgety mode''~\cite{FidgetyAFL}, while \mopt is specified with \verb|-L 0| to facilitate its ``pacemaker mode''~\cite{mopt}. The CPU affinity is turned off during fuzzing to avoid multiple threads being bound to a single CPU core.
During fuzzing, we run each of the aforementioned fuzzers \emph{six times} against all the 12 benchmark programs, with a time budget of 24 hours. Since all the evaluated programs are set to run with four working threads and the threads are mapped to different cores, it takes \emph{each fuzzer} approximately $12\times 6\times 24\times 4=6912$ CPU hours. 

\begin{table*}[t]
\setlength{\tabcolsep}{2.2pt} 
\caption{Fuzzing results on \muzz, \mafl, AFL and \mopt, in terms of generated seeds. $\testsALL$: total number of new seeds; $\testsMT$: number of new multithreading-relevant seeds; \testsRatio: the percentage of multithreading-relevant seeds among all the generated seeds. \textbf{Bold data entries} mark the best results among the fuzzers, in terms of $\testsMT$ and \testsRatio. The numbers in parentheses (for $\testsALL$ and \testsRatio) denote the differences between \muzz and the others; for example, ``(+1850)'' is the more multithreading-relevant seeds generated by \muzz: 5127 than \mafl: 3277.}
\label{tbl:eval_seeds}
\centering
\begingroup
\renewcommand{\arraystretch}{0.9} 
\footnotesize
\begin{tabular}{c A r B r C r A r B r C r A r B r C r A r B r C r}
\toprule
\multirow[c]{3}{*}{\textbf{ID}} & \multicolumn{3}{c}{\textbf{\muzz}} & \multicolumn{3}{c}{\textbf{\mafl}} & \multicolumn{3}{c}{\textbf{AFL}} & \multicolumn{3}{c}{\textbf{\mopt}} \\ 
\cmidrule(lr){2-4} \cmidrule(lr){5-7} \cmidrule(lr){8-10} \cmidrule(lr){11-13}
& \textbf{\testsALL} & \textbf{\testsMT} & \textbf{\testsRatio} 
& \textbf{\testsALL} & \textbf{\testsMT} & \textbf{\testsRatio} 
& \textbf{\testsALL} & \textbf{\testsMT} & \textbf{\testsRatio}
& \textbf{\testsALL} & \textbf{\testsMT} & \textbf{\testsRatio}
\\ 
\midrule
\textbf{lbzip2-c} 
& {8056} & \gres{5127} & \gres{63.6\%} 
&  6307   &  3277\mydiff{+1850}  &  52.0\%\mydiff{+11.7\%} 
&  5743   & 2464\mydiff{+2663}   &  42.9\%\mydiff{+20.7\%} 
&  6033   & 2524\mydiff{+2603}   &  41.8\%\mydiff{+21.8\%}
\\ 
\textbf{pbzip2-c} 
& {381} & \gres{126} & \gres{33.1\%}
&  340    &  91\mydiff{+35}    &  26.8\%\mydiff{+6.3\%}  
&  272    & 69\mydiff{+57}    &  25.4\%\mydiff{+7.7\%}  
&  279    & 71\mydiff{+55}    &  25.4\%\mydiff{+7.6\%}
\\ 
\textbf{pbzip2-d} 
& {1997} & \gres{297} & \gres{14.9\%} 
&  1706    &  119\mydiff{+178}    &  7.0\%\mydiff{+7.9\%}   
&  1650    & 68\mydiff{+229}     &  4.1\%\mydiff{+10.8\%}   
&  1623    & 62\mydiff{+235}     &  3.8\%\mydiff{+11.1\%}
\\ 
\textbf{pigz-c}   
& {1406} & \gres{1295} & \gres{92.1\%} 
&  1355    &  1189\mydiff{+106}   &  87.7\%\mydiff{+4.4\%}  
&  1298    & 1098\mydiff{+197}   &  84.6\%\mydiff{+7.5\%}  
&  1176       & 982\mydiff{+313}       & 83.5\%\mydiff{+8.6\%}
\\ 
\textbf{pxz-c}    
& {7590} & \gres{5249} & \gres{69.2\%} 
&  5637   &  3401\mydiff{+1848}  &  60.3\% \mydiff{+8.8\%} 
&  5357   & 2470\mydiff{+2779}   &  46.1\% \mydiff{+23.0\%} 
&  5576       & 2634\mydiff{+2615}       &  47.2\% \mydiff{+21.9\%}
\\ 
\textbf{xz-c}     
& {2580} & \gres{1098} & \gres{42.6\%} 
&  2234   &  767\mydiff{+331}   &  34.3\%\mydiff{+8.2\%}  
&  1953   & 581\mydiff{+517}   &  29.7\%\mydiff{+12.8\%}  
& 1845        & 566\mydiff{+532}       & 30.7\%\mydiff{+11.9\%}
\\ 
\textbf{gm-cnvt}  
& {15333} & \gres{13774} & \gres{89.8\%} 
&  14031   &  10784\mydiff{+2990}  &  76.9\%\mydiff{+13.0\%}  
&  12453   & 8290\mydiff{+5484}  &  66.6\%\mydiff{+23.3\%}  
&  12873       &    8956\mydiff{+4818}    & 69.6\%\mydiff{20.3\%}
\\ 
\textbf{im-cnvt}  
& {14377} & \gres{12987} & \gres{90.3\%} 
&  12904   &  10610\mydiff{+2377}  &  82.2\%\mydiff{+8.1\%}  
&  9935   & 7634\mydiff{+5353}  &  76.8\%\mydiff{+76.8\%}  
&  10203       &    8012\mydiff{+4975}    & 78.5\%\mydiff{+11.8\%}
\\ 
\textbf{cwebp}    
& {11383} & \gres{7554} & \gres{66.4\%} 
&  10389   &  6868\mydiff{+686}  &  66.1\% \mydiff{+0.3\%} 
&  9754   & 5874\mydiff{+1680}  &  60.2\% \mydiff{+6.1\%} 
&  9803       & 5869\mydiff{+1685}       & 59.9\%\mydiff{+6.5\%}
\\ 
\textbf{vpxdec}   
& {28892} & \gres{25593} & \gres{88.6\%} 
&  27735  &  22507\mydiff{+3086}  &  81.2\%\mydiff{+7.4\%}  
&  24397  & 18936\mydiff{+6657}  &  77.6\%\mydiff{+11.0\%}  
&  27119       &    20896\mydiff{+4697}    & 77.1\%\mydiff{11.5\%}
\\ 
\textbf{x264}     
& {15138} & \gres{14611} & \gres{96.5\%} 
&  14672   &  13413\mydiff{+1198}  &  91.4\% \mydiff{+5.1\%} 
&  13211   & 11801\mydiff{+2810}  &  89.3\%  \mydiff{+7.2\%}
&  12427       &    11202\mydiff{+3409}    & 90.1\%\mydiff{+6.4\%}
\\ 
\textbf{x265} 
& 12965  & {10704}  & \gres{82.6\%}  
&  {13858} & \gres{10890}\mydiff{-186}  &  78.6\% \mydiff{+4.0\%} 
&  12980   & 9957\mydiff{+747}  &  76.7\% \mydiff{+5.9\%} 
&  13142       &    10154 \mydiff{+550}   & 77.3\%\mydiff{+5.3\%}
\\ 
\bottomrule
\end{tabular}
\endgroup
\end{table*}
 
\subsection{Seed Generation (RQ1)}\label{sec:rq_seeds}

Table~\ref{tbl:eval_seeds} shows the overall fuzzing results 
in terms of {newly generated seeds}. We collect the total number of generated seeds (\testsALL) and the number of seeds that exercise the multithreading context (\testsMT). In AFL's jargon, {\testsALL} corresponds to the {distinct paths} that the fuzzer observes~\cite{afl_detail}. The multithreading-relevant seeds are collected with a separate procedure, based on the observations that they at least invoke one element in \TStart. Therefore, \testsMT tracks the different multithreading execution states during fuzzing --- a larger value of this metric suggests the fuzzer can keep more effective thread-interleaving seeds.
We sum up those seed files across all six fuzzing runs to form 
\testsALL~ and \testsMT~in Table~\ref{tbl:eval_seeds}. The {\testsRatio} column shows the percentage of {\testsMT} over \testsALL. \testsRatio determines the probability of picking a multithreading-relevant seed during seed selection, which greatly impacts the overall quality of the generated seeds. 
Obviously, the most critical metrics are \testsMT and \testsRatio.

\muzz surpasses \mafl, AFL, and \mopt in both metrics.
First, \muzz exhibits superiority in generating multithreading-relevant seeds --- in all the benchmarks \muzz achieves the highest \testsMT. 
For example, in \projfont{pbzip2-d}, despite that all the {\testsRatio} are relatively small, \muzz generated 297 multithreading-relevant seeds, which is 178 more than \mafl (119), 229 more than AFL (68), and 235 more than \mopt (62). Moreover, for larger programs such as \projfont{im-cnvt} (binary size 19.4M), $\testsMT$ of \muzz (12987) is still better than the others (\mafl: 10610, AFL: 7634, \mopt: 8012).
Second, the value of \testsRatio in \muzz is more impressive 
--- \muzz wins the comparison over all the benchmarks. For example, in  \projfont{pbzip2-d}, 
\muzz's result of \testsRatio is higher --- \muzz: 14.9\%, AFL: 7.0\% \mafl: 4.1\%, and \mopt: 3.8\%. 
For the benchmark where AFL has already achieved a decent result, e.g., 89.3\% for \projfont{x264}, \muzz 
can even improve it to 96.5\%. Meanwhile, although \mafl has the largest {\testsMT} for \projfont{x265} (10890), the value of its \testsRatio (78.6\%) is less than that of \muzz (82.6\%).

It is worth noting that \mafl also outperforms AFL and \mopt w.r.t. {\testsMT} and {\testsRatio} in all the benchmarks. For example, in \projfont{pxz-c}, the number of generated multithreading-relevant seeds in \mafl is 3401, which is more than AFL (2470) and \mopt (2634). Correspondingly, the percentage of multithreading-relevant seeds in \mafl is 60.3\%;  for AFL and \mopt, they are 46.1\% and 47.2\%, respectively. 
Considering \mafl, AFL, \mopt apply coverage-oriented instrumentation (\MTIns), we can conclude that other strategies in \mafl, including thread-context instrumentation, schedule-intervention instrumentation, and the optimized dynamic strategies, also contribute to effective seed generation.

\begin{tcolorbox}[size=title]
{\textbf{Answer to RQ1: } \muzz has advantages in increasing the number and percentages of multithreading-relevant seeds for multithreaded programs. The proposed three thread-aware instrumentations and dynamic fuzzing strategies benefit the seed generation.}
\end{tcolorbox}

\subsection{Vulnerability Detection (RQ2)}\label{sec:rq_Vuls}

For vulnerability detection, we denote the total number of proof-of-crash (POC) files generated during fuzzing as \crashesNUM. The vulnerability analysis component (right-bottom area as \PvulS in Figure~\ref{fig:overview}) analyzes the POC files and categorizes them into different vulnerabilities. This basically follows ClusterFuzz's practice~\cite{clusterfuzz}: if two POC files have the same last N lines of backtraces and the root cause is the same (e.g., both exhibit as \emph{buffer-overflow}), they are treated as one vulnerability. Afterwards, we manually triage all 
the vulnerabilities into two groups based on their relevance with multithreading: the concurrency-vulnerabilities \VulsMT, and the other vulnerabilities that do not occur in multithreading context \VulsST. The number of these vulnerabilities are denoted as {\NVulsMT} and {\NVulsST}, respectively.

\begin{table*}[t]
\setlength{\tabcolsep}{4.1pt} 
\caption{Fuzzing results on \muzz, \mafl, AFL and \mopt, in terms of crashes and vulnerabilities. Some projects (e.g., \projfont{lbzip2-c}) are excluded since there were no crashes/vulnerabilities detected by any of the fuzzers. $\crashesNUM$: number of proof-of-crash (POC) files; $\crashesMT$: number of multithreading-relevant POC files; $\NVulsMT$: number of concurrency-vulnerabilities. \crashesST: number of POC files irrelevant with multithreading; $\NVulsST$: number of vulnerabilities irrelevant to multithreading. \textbf{Bold data entries} mark the best results for \crashesMT and \NVulsMT. The numbers in parentheses denote the differences between \muzz and others.}
\label{tbl:eval_vuls}
\centering
\begingroup
\renewcommand{\arraystretch}{0.9} 
\footnotesize
\begin{tabular}{c ArBrCrDrEr ArBrCrErEr ArBrCrDrEr ArBrCrDrEr}
\toprule
\multirow[c]{3}{*}{\textbf{ID}} & \multicolumn{5}{c}{\textbf{\muzz}} & \multicolumn{5}{c}{\textbf{\mafl}} & \multicolumn{5}{c}{\textbf{AFL}} & \multicolumn{5}{c}{\textbf{\mopt}} \\ 
\cmidrule(lr){2-6} \cmidrule(lr){7-11} \cmidrule(lr){12-16} \cmidrule(lr){17-21}
& \textbf{\crashesNUM} & \textbf{\crashesMT} & \textbf{\NVulsMT} & \textbf{\crashesST} & \textbf{\NVulsST}  
& \textbf{\crashesNUM} & \textbf{\crashesMT} & \textbf{\NVulsMT} & \textbf{\crashesST} & \textbf{\NVulsST}
& \textbf{\crashesNUM} & \textbf{\crashesMT} & \textbf{\NVulsMT} & \textbf{\crashesST} & \textbf{\NVulsST} 
& \textbf{\crashesNUM} & \textbf{\crashesMT} & \textbf{\NVulsMT} & \textbf{\crashesST} & \textbf{\NVulsST}
\\ 
\midrule
\textbf{pbzip2-c}
& {6}  & \gres{6}  &   \gres{1} & {0}  &  0
& {6} & {0}\mydiff{+6}  & \gres{1}\mydiff{0} & {0}\mydiffn{0} & 0\mydiffn{0}  
&  0 & {0}\mydiff{+6}  & 0\mydiff{+1} & {0}\mydiffn{0} & 0\mydiffn{0}          
&  0 & {0}\mydiff{+6}  & 0\mydiff{+1} & {0}\mydiffn{} & 0\mydiffn{0}
\\ 
\textbf{pbzip2-d} 
&  {15} & \gres{15}  &   \gres{1} & {0}  & 0      
& 0 & {0}\mydiff{+15}  & 0\mydiff{+1} & {0}\mydiffn{0} & 0\mydiffn{0}                 
&  0 & {0}\mydiff{+15}  & 0\mydiff{+1} & {0}\mydiffn{0} & 0\mydiffn{0}        
&  0 & {0}\mydiff{+15}  & 0\mydiff{+1} & {0}\mydiffn{0} & 0\mydiffn{0}
\\ 
\textbf{im-cnvt} 
& {87} & \gres{63} &  \gres{4} & {24} &  1   
&  49 & {23}\mydiff{+40} & 2\mydiff{+2} & {26}\mydiffn{-2} & 1\mydiffn{0}              
&  29 & {6}\mydiff{+57} & 2\mydiff{+2} & {23}\mydiffn{+1} & 1\mydiffn{0}         
&  32 & {6}\mydiff{+57} & 2\mydiff{+2} & {26}\mydiffn{-2} & 1\mydiffn{0}
\\ 
\textbf{cwebp}    
&  19   & {0}  &   0 & {19}  &   1                 
& {27} & {0}\mydiff{0} & 0\mydiff{0} & {27}\mydiffn{-8} & 1\mydiffn{0}          
&  14 & {0}\mydiff{0}  & 0\mydiff{0} & {14}\mydiffn{+5} & 1\mydiffn{0}          
&  15 & {0}\mydiff{0}  & 0\mydiff{0} & {15}\mydiffn{+4} & 1\mydiffn{0}
\\ 
\textbf{vpxdec}   
& {523} & \gres{347}  &   \gres{2}  & {176}  &   2    
& {495} & {279}\mydiff{+68} & 1\mydiff{+1} & {216}\mydiffn{-30} & 2\mydiffn{0}          
&  393 & {205}\mydiff{+142} & 1\mydiff{+1} & {188}\mydiffn{-12} & 2\mydiffn{0}          
&  501  & {301}\mydiff{+46} & 1\mydiff{+1} & {200}\mydiffn{-24} & 2\mydiffn{0}
\\ 
\textbf{x264}     
& {103} & \gres{103}   &   \gres{1}  & {0}  &   0   
& 88 & {88}\mydiff{+15}  & \gres{1}\mydiff{0} & {0}\mydiffn{0} & 0\mydiffn{0}                 
&  78 & {78}\mydiff{+25}  & \gres{1}\mydiff{0} & {0}\mydiffn{0} & 0\mydiffn{0}          
&  66 & {66}\mydiff{+37}  & \gres{1}\mydiff{0} & {0}\mydiffn{0} & 0\mydiffn{0}
\\ 
\textbf{x265} 
&  43 & {0}   &   0 & {43}  &   1                           
& 52 & {0}\mydiff{0}  & 0\mydiff{0} & {52}\mydiffn{-9} & 1\mydiffn{0}      
&  {62} & {0}\mydiff{0}  & 0\mydiff{0} & {62}\mydiffn{-19} & 1\mydiffn{0}   
&  59 & {0}\mydiff{0}  &  0\mydiff{0} & {59}\mydiffn{-16}     & 1\mydiffn{0}
\\ 
\bottomrule
\end{tabular}
\endgroup
\end{table*}
 
We mainly refer to \crashesMT, \NVulsMT in Table~\ref{tbl:eval_vuls} to evaluate \muzz's \emph{concurrency-vulnerability} detection capability.

The number of multithreading-relevant POC files, \crashesMT, is important since it corresponds to different crashing states when executing multithreading context ~\cite{fuzz_survey,ccs18_eval_fuzzing}. It is apparent that \muzz has the best results of \crashesMT in all the benchmarks that have \VulsMT vulnerabilities (e.g., for \projfont{im-cnvt}, \muzz: 63, \mafl: 23, AFL: 6, \mopt: 6). Moreover, \mafl also exhibits better results than AFL and \mopt (e.g., for \projfont{pbzip2-c}, \muzz: 6, \mafl: 6, AFL: 0, \mopt: 0). This suggests that \muzz's and \mafl's emphasis on multithreading-relevant seed generation indeed helps to exercise more erroneous multithreading-relevant execution states.

The most important metric is {\NVulsMT} since our focus is to detect concurrency-vulnerabilities 
(\VulsMT). Table \ref{tbl:eval_vuls} shows that \muzz has the best results: \muzz detects 9 concurrency-vulnerabilities, while \mafl, AFL and \mopt detects 5, 4, 4, respectively.
Detected \VulsMT can be divided into three groups.

\textbf{1) \VulsMT caused by concurrency-bugs}.
We term this group of vulnerabilities as \texorpdfstring{\VulsCB}{}.
The 4 vulnerabilities in \projfont{im-cnvt} all belong to this group --- the misuses of caches shared among threads cause the data races. The generated seeds may exhibit various symptoms such as \emph{buffer-overflow} and \emph{memcpy-param-overlap}. \muzz found all the 4 vulnerabilities, while the others only found 2. We also observed that for the 2 vulnerabilities that are detected by all these fuzzers, \mafl's detection capability appears \emph{more stable} since it detects both in all its six fuzzing runs, while the others can only detect them at most in five runs (not depicted in the table). 
\textbf{2) \texorpdfstring{\VulsMT}{} triggered in multithreading only but not induced by concurrency-bugs}.
For example, the vulnerability in \projfont{pbzip2-d} stems from a \emph{stack-overflow} error when executing a function concurrently.
This crash can never happen when \projfont{pbzip2-d} works in single-thread mode since it does not even invoke that erroneous function. In our evaluation, \muzz detected this vulnerability while the other fuzzers failed. Another case is the vulnerability in \projfont{pbzip2-c}, which was detected by \muzz and \mafl, but not by AFL or \mopt.
3) \textbf{Other concurrency-vulnerabilities}. The characteristics of these \VulsMT are that their crashing backtrace contains multithreading context (i.e., \TStart is invoked), however, the crashing condition might also be occasionally triggered when only one thread is specified.
The {\VulsMT} vulnerabilities detected in \projfont{vpxdec} and \projfont{x264} belong to this category.
In particular, \muzz detects 2 vulnerabilities in \projfont{vpxdec} while \mafl, AFL, and \mopt only find 1.

We consider the reason behind the differences w.r.t. \crashesMT and \NVulsMT among the fuzzers to be that, \muzz keeps more ``deeper'' multithreading-relevant seeds that witness different execution states, and mutations on some of them are more prone to trigger the crashing conditions.

Columns \crashesNUM, \crashesST, \NVulsST are metrics less focused. But we can still observe that 1) according to \crashesNUM, \muzz (and \mafl) can exercise more total crashing states; 2) despite that the values of \crashesST from \muzz are usually smaller, \muzz can still find all the (categorized) \VulsST detected by other fuzzers.

From the 12 evaluated benchmarks, we reported the 10 new vulnerabilities (sum of \muzz's results in columns {\NVulsMT} and {\NVulsST} except for row \projfont{vpxdec}; 7 of them belong to \VulsMT), all of them have been confirmed or fixed, 3 of which have already been assigned CVE IDs.
Besides, we also conducted a similar evaluation on \projfont{libvpx} v1.8.0-178 (the git HEAD version at the time of evaluation). \muzz detected a 0-day concurrency-vulnerability within 24 hours (among six fuzzing runs, two of them detected the vulnerability in 5h38min and 16h07min, respectively), while \mafl, AFL and \mopt failed to detect it in 15 days (360 hours) in all their six fuzzing runs. 
The newly detected vulnerability has been assigned with another CVE ID. The vulnerability details are available in Table~\ref{tbl:bugs}.

Given the fact that there are extremely few CVE records caused by concurrency-vulnerabilities (e.g., 202 among 70438, based on records from CVE-2014-* to CVE-2018-*)~\cite{cve-db}, \muzz demonstrates the high capability in detecting concurrency-vulnerabilities.

\begin{tcolorbox}[size=title]
{\textbf{Answer to RQ2: } \muzz demonstrates {superiority} in exercising more multithreading-relevant crashing states and detecting concurrency-vulnerabilities.}
\end{tcolorbox}

\begin{table*}[t]
\caption{Comparisons of replay patterns \replayRO and \replayRN on \muzz, \mafl, AFL and \mopt, in terms of concurrency violations (\replayCN) and concurrency-bugs (\replayCB). The best results of {\replayCN} and {\replayCB} are \bres{underlined} / \gres{bold} for {\replayRO} / {\replayRN} respectively.}
\label{tbl:eval_BugsMT}
\centering
\begingroup
\setlength{\tabcolsep}{6.0pt} 
\renewcommand{\arraystretch}{1.0} 
\footnotesize
\begin{tabular}{c BrCr BrCr BrCr BrCr ?{1.2mm} BrCr BrCr BrCr BrCr}
\toprule
\multirow[c]{3}{*}{\textbf{ID}}
 
& \multicolumn{8}{c}{\textbf{\replayRO}} & \multicolumn{8}{c}{\textbf{\replayRN}} \\ 
\cmidrule(lr){2-9} \cmidrule(lr){10-17}
 & \multicolumn{2}{c}{\textbf{\muzz}} & \multicolumn{2}{c}{\textbf{\mafl}} & \multicolumn{2}{c}{\textbf{AFL}} & \multicolumn{2}{c}{\textbf{\mopt}} & \multicolumn{2}{c}{\textbf{\muzz}} &  \multicolumn{2}{c}{\textbf{\mafl}} & \multicolumn{2}{c}{\textbf{AFL}} & \multicolumn{2}{c}{\textbf{\mopt}} \\

\midrule

 & \textbf{\replayCN} & \textbf{\replayCB} & \textbf{\replayCN} & \textbf{\replayCB} 
 & \textbf{\replayCN} & \textbf{\replayCB} & \textbf{\replayCN} & \textbf{\replayCB} 
 & \textbf{\replayCN} & \textbf{\replayCB} & \textbf{\replayCN} & \textbf{\replayCB}
 & \textbf{\replayCN} & \textbf{\replayCB} & \textbf{\replayCN} & \textbf{\replayCB} 
 \vspace{0.1mm}
 \\
\textbf{lbzip2-c} & \bres{469} & \bres{1} & 447 & \bres{1} & {386} & \bres{1} & 435 & \bres{1} 
& \gres{493} & \gres{1} & 483 & \gres{1} & 421 & \gres{1} & 458 & \gres{1}\\ 
\textbf{pigz-c} & 793 & \bres{1} & \bres{803} & \bres{1} & 764 & \bres{1} & 789 & \bres{1} 
& 856 & \gres{1} & \gres{862} & \gres{1} & 727 & \gres{1} & 742 & \gres{1} \\ 
\textbf{gm-cnvt} & \bres{93} & \bres{5} & 79 & 4 & 45 & 2 & 55 & 3 
& \gres{133} & \gres{5} & {83} & 4 & 54 & 3 & 57 & 3\\ 
\textbf{im-cnvt} & \bres{92} & \bres{3}  & 84 & \bres{3} & 58 & 2 & 56 & 2 
& \gres{118} & \gres{3} & 117 & \gres{3} & 65 & 2 & 59 & 2\\ 
\textbf{vpxdec} & \bres{31} & \bres{3} & 17 & 1 & 23 & 1 & 18 & 1 
& \gres{42} & \gres{3} & 22 & 1 & 25 & 1 & 22 & 1\\ 
\textbf{x264} & \bres{68} & \bres{8} & 46 & 6 & 28 & 4 & 30 & 5 
& \gres{91} & \gres{9} & 52 & 6 & 25 & 4 & 28 & 4\\ 
\bottomrule
\end{tabular}
\endgroup
\end{table*}
 
\subsection{Concurrency-bug Revealing (RQ3)}\label{sec:rq_BugsMT}

The fuzzing phase only detects the vulnerabilities caused by crashes, but the \emph{seemingly normal} seed files generated during fuzzing may still execute paths that trigger concurrency-violation conditions like \emph{data-races}, \emph{deadlocks}, etc. We detect concurrency-bugs in {concurrency-bug revealing} component (\PcbugS, right-top in Figure~\ref{fig:overview}). It is worth noting that our goal is \emph{not} to improve the capabilities of concurrency-bug detection over existing techniques such as \ts~\cite{kcc:tsan}, Helgrind~\cite{helgrind}, or UFO~\cite{icse18_ufo}. Instead, we aim to \emph{reveal as many bugs as possible within a time budget, by replaying against fuzzer-generated seeds with the help of these techniques}. In practice, this component feeds the target program with the seeds that were generated during fuzzing as its inputs, and facilitate detectors such as \ts to reveal concurrency-bugs.
During this evaluation, we compiled the target programs 
with \ts and replayed them against the fuzzer-generated multithreading-relevant seeds (corresponding to {\testsMT} in Table~\ref{tbl:eval_seeds}). We did not replay with \emph{all the generated seeds} (corresponding to {\testsALL} in Table~\ref{tbl:eval_seeds}) since seeds not exercising multithreading context will not reveal concurrency-bugs.

We limit our replay time budget to two hours; in { ~\S\ref{sec:app-replay-budget} we discuss the rationale of this configuration}. The next is to \emph{determine the replay pattern} per seed to reveal more concurrency-bugs within this budget. This is necessary since \ts may fail to detect concurrency-bugs in a few runs when it does not \emph{observe concurrency violation conditions}~\cite{pldi09_fasttrack,helgrind,kcc:tsan}.
Meanwhile, as the time budget is limited, we cannot exhaustively replay against a given seed to see whether it may trigger concurrency-violations --- in the worst case, we may waste time in executing against a seed that never violates the conditions. We provide two replay patterns.
\begin{enumerate}[labelindent=4pt]
	\item[\replayRO] It executes against each seed in the queue \emph{once per turn} in a round-robin way, until reaching the time budget.
	\item[\replayRN] It relies on \Ncal in repeated execution (c.f., \S\ref{sec:cal_exec}): each seed is executed \emph{$\frac{\Ncal}{\NcalO}$ times per turn} continuously in a round-robin way. According to Equation~\ref{eq:afl_cal}, we replay 5 times per turn (40/8) for AFL generated multithreading-relevant seeds; for \muzz and \mafl, it is determined by Equation~(\ref{eq:mtfuzz_cal}), with candidate values 2, 3, 4, 5. 
\end{enumerate}

It is fair to compare replay results w.r.t. \replayRO and \replayRN in that the time budget is fixed. The difference between the two patterns is that seeds' execution orders and accumulated execution time spent on them can be rather different.

Table~\ref{tbl:eval_BugsMT} depicts the results for concurrency-bug revealing with \replayRO and \replayRN.
{\replayCN} is the number of \emph{observed concurrency-violation executions} and
{\replayCB} is the number of {concurrency-bugs (\BugsMT)} according to their root causes. For example, it only counts one concurrency-bug (\replayCB=1) even when the replaying process observes 10 data-race pairs across executions (\replayCN=10), as long as the root cause of the races is unique. 
We analyze this table from two perspectives.

\emph{First, \muzz demonstrates superiority in concurrency-bug detection regardless of replay patterns.} This is observed based on the ``best results'' for each metric in each pattern. \muzz achieves the best results for most projects. For example, when \projfont{x264} is replayed with \replayCN, 1) \muzz's found the most violations --- the values of {\replayCN} are, \muzz: 68, \mafl: 46, AFL: 28, \mopt: 30; 2) the best result of {\replayCB} also comes from \muzz, \muzz: 8, \mafl: 6, AFL: 4, \mopt: 5. Similar results can also be observed with \replayRN for \projfont{x264}, where \muzz has the biggest \replayCN (91) and biggest \replayCB (9). The only project where \mafl achieves the best is \projfont{pigz-c}, where it is slightly better than \muzz.

\emph{Second, as to \muzz and \mafl, \replayRN is probably better than \replayRO.} It is concluded based on the fact that \replayRN's ``best results'' are all better than \replayRO's.
For example, as to {\replayCN} in \projfont{x264}, the best result of {\replayCN} is achieved with \replayRN (\replayRO: 68, \replayRN: 91); similarly, the best result of {\replayCB} also comes from \replayRN (\replayRO: 8, \replayRN: 9).
In the meantime, there seems to be no such implication inside AFL or \mopt. Besides the numbers of concurrency-violations or concurrency-bugs, \S\ref{sec:app-cb-tte} provides a case study on \codefont{gm-cnvt} that demonstrates \replayRN's advantages over \replayRO w.r.t. \emph{time-to-exposure} of the concurrency-bugs.

We have reported all the newly detected 19 concurrency-bugs (excluding the 3 {concurrency-bugs} in \projfont{vpxdec-v1.3.0-5589}) to their project maintainers (c.f., Table~\ref{tbl:bugs} for the details).

\begin{tcolorbox}[size=title]
{\textbf{Answer to RQ3: } 
{\muzz} {outperforms} competitors in detecting concurrency-bugs; the value \Ncal calculated during fuzzing additionally contributes to revealing these bugs.}
\end{tcolorbox}

\begin{table*}[t]
    \caption{Newly detected vulnerabilities and concurrency-bugs. This summarizes the \emph{new} vulnerabilities and concurrency-bugs evaluated in Table~\ref{tbl:eval_vuls} and Table~\ref{tbl:eval_BugsMT} over the 11 benchmarks (\projfont{libvpx-v1.3.0-5589} results are all excluded), and includes one concurrency-vulnerability in \projfont{vpxdec-v1.8.0-178} which was mentioned in \S\ref{sec:rq_Vuls}.}
    \label{tbl:bugs}
    \centering\small
    \begingroup
    \setlength{\tabcolsep}{4.8pt} 
    \renewcommand{\arraystretch}{0.86} 
    \begin{tabular}{ccccccccc}
        \toprule
        \textbf{Bugs} & \textbf{Project} & \textbf{Bug Type} & \textbf{Reported Category} & \textbf{\muzz} & \textbf{\mafl} & \textbf{AFL} &\textbf{\mopt} & \textbf{Status} \\
        \midrule
        \textbf{V1} & pbzip2 & \VulsMT & divide-by-zero &\yyy&\yyy&\xxx& \xxx& confirmed, not fixed \\
        \textbf{V2} & pbzip2 & \VulsMT & stack-overflow &\yyy&\xxx&\xxx& \xxx& confirmed, not fixed \\
        \textbf{V3} & ImageMagick & \VulsMT & memcpy-param-overlap &\yyy&\xxx&\xxx& \xxx& CVE-2018-14560 \\
        \textbf{V4} & ImageMagick & \VulsMT & buffer-overflow &\yyy&\yyy&\yyy& \yyy& CVE-2019-15141 \\
        \textbf{V5} & ImageMagick & \VulsMT & buffer-overflow &\yyy&\yyy&\yyy& \yyy& confirmed, fixed \\
        \textbf{V6} & ImageMagick & \VulsMT & buffer-overflow &\yyy&\xxx&\xxx&\xxx& confirmed, fixed \\
        \textbf{V7} & ImageMagick & \VulsST & negative-size-param &\yyy&\yyy&\yyy&\yyy& CVE-2018-14561\\
        \textbf{V8} & x264 & \VulsMT & buffer-overflow &\yyy&\yyy&\yyy&\yyy& confirmed, fixed \\
        \textbf{V9} & libwebp & \VulsST & failed-to-allocate &\yyy&\yyy&\yyy&\yyy& confirmed, won't fix \\
        \textbf{V10} & x265 & \VulsST & divide-by-zero &\yyy&\yyy&\yyy&\yyy& confirmed, not fixed \\
        \textbf{V11} & libvpx-v1.8.0-178 & \VulsMT & invalid-memory-read &\yyy&\xxx&\xxx&\xxx & CVE-2019-11475\\
        \midrule
        \textbf{C1} & lbzip2 & \BugsMT & thread-leak &\yyy&\yyy&\yyy&\yyy& confirmed, not fixed \\
        \textbf{C2} & pigz & \BugsMT & lock-order-inversion &\yyy&\yyy&\yyy&\yyy& confirmed, fixed \\
        \textbf{C3-C7} & GraphicsMagick & \BugsMT & data-race & 5 & 4 & 3 & 2 &  confirmed, fixed \\
        \textbf{C8-C10} & ImageMagick & \BugsMT & data-race & 3 & 3 & 2 & 2 & confirmed, fixed \\
        \textbf{C11-C19} & x264 & \BugsMT & data-race & 9 & 6 & 4 & 4 & confirmed, not fixed\\
        \bottomrule
    \end{tabular}
    \endgroup
\end{table*}
 
\subsection{Further Discussions}\label{sec:discuss}
This section discusses miscellaneous concerns, issues and observations for \muzz's design and evaluation.

\subsubsection{Constant Parameters}\label{sec:app-const-params}
Using empirical constant parameters for grey-box fuzzing is practiced by many fuzzing techniques~\cite{afl_detail,mopt,hawkeye}. For example, AFL itself has many hard-coded configurations 
used by default; \mopt additionally has the suggested configuration to control the time to move on to pacemaker mode (i.e., \verb|-L 0|).

In \muzz, constant parameters are used in two places.

\textbf{(1) The upper bounds for coverage-oriented instruction: \stpnO (defaults to 0.5) and \mtpnO (defaults to 0.33)}.
These default values inspire from AFL's ``selective deputy instruction instrumentation'' strategy to make the instrumentation ratio to be 0.33 when AddressSanitizer is involved during instrumentation
.
Larger values of \stpnO and \mtpnO increases the instrumentation ratio only if \emph{the thresholds are frequently reached}. Subsequently, the instrumented program has these symptoms:
a) the program size after instrumentation increases; b) the execution state feedback is potentially better; c) the instrumentation-introduced execution speed slowdown is more evident.
Therefore, increasing the values of \stpnO and \mtpnO reflects a tradeoff between precise feedback and its overhead.
In our benchmarks, when we assign \stpnO=0.5, \mtpnO=0.33,
\begin{enumerate}[label=$\bullet$,noitemsep,topsep=0pt,parsep=0pt,labelindent=0pt,labelwidth=*]
    \item For \projfont{im-cnvt}, the speed slowdown is about 15\% compared to default settings, while the capability of detecting concurrency-vulnerabilities and concurrency-bugs are similar; meanwhile, there are a few more multithreading-relevant seeds (\testsMT) but \testsRatio is slightly smaller.
    \item For \projfont{pbzip2-c}, the differences brought by changes of \stpnO and \mtpnO from the default settings are all neglectable.
\end{enumerate}
We believe there are no optimal instrumentation thresholds that work for all the projects; therefore \muzz provides the empirical values as the defaults.

\textbf{(2) The seed selection probabilities $\probYNT=0.95$, $\probYNN=0.01$, $\probNNN=0.15$ in Algorithm~\ref{algo:select_seed}.}
These constants are not introduced by \muzz, but based on AFL's ``skipping probability'' to conditionally favor seeds with new coverage~\cite{afl_detail}. 

Since the 12 benchmarks that we chose are quite diversified (c.f., \S\ref{sec:bm_stats}), it is considered fair to use default settings for these parameters, when comparing \muzz, \mafl with other fuzzers such as AFL, \mopt. In practice, we suggest keeping \muzz's default settings to test other multithreaded programs.

\subsubsection{Schedule-intervention Instrumentation}\label{sec:app-schedule-intervention}
The goal of \muzz's schedule-intervention is to diversify interleavings during repeated executions in the fuzzing phase.
During the evaluation, we did not separately evaluate the effects of \emph{schedule-intervention instrumentation}. However, based on our observation, this instrumentation is important to achieve more stable fuzzing results. Two case studies can support this statement.
\begin{enumerate}[label=\alph*),noitemsep,topsep=0pt,parsep=0pt,labelindent=0pt,labelwidth=*]
    \item We turned off schedule-intervention instrumentation in \muzz and fuzzed \projfont{lbzip2-c} six times on the \emph{same} machine. The calculated value of \testsRatio is 54.5\% (= 4533/8310), which is lower than the result in Table~\ref{tbl:eval_seeds} (63.6\% = 5127/8056). Since 54.5\% is still greater than the results of AFL (42.9\%) and \mopt (41.8\%), this also indicates \muzz's other two strategies indeed benefit the multithreading-relevant seed generation for fuzzing.
    \item We turned off schedule-intervention instrumentation in \muzz and fuzzed \projfont{im-cnvt} on a \emph{different} machine. In all the six fuzzing runs it only detects three concurrency-vulnerabilities which is less than the result in Table~\ref{tbl:eval_vuls} (\NVulsMT=4). Meanwhile, when the schedule-intervention instrumentation is re-enabled, \muzz can still detect four concurrency-vulnerabilities in that machine. 
\end{enumerate}

\subsubsection{Time-to-exposure for Concurrency-bug Revealing}\label{sec:app-cb-tte}
In \S\ref{sec:rq_BugsMT}, we demonstrate \replayRN's advantage over \replayRO in terms of occurrences of concurrency-violations ({\replayCN}) and the number of categorized concurrency-bugs ({\replayCB}). Another interesting metric is the \emph{time-to-exposure} capability of these two replay patterns --- given the ground truth that the target programs contain certain concurrency-bugs, the minimal time cost for each pattern to reveal all the known bugs. This metric can further distinguish the two replay patterns' capabilities in terms of revealing concurrency-bug.

We conducted a case study on \projfont{gm-cnvt}. From Table~\ref{tbl:eval_BugsMT}, it is observable that with both \replayRO and \replayRN, \ts detected four concurrency-bugs (\replayCB) by replaying the \mafl generated multithreading-relevant seeds (totally 10784) from Table~\ref{tbl:eval_seeds}; besides, their {\replayCN} results are also similar (\replayRO: 79, \replayRN: 83). 
We repeated six times against the 10784 seeds by applying \replayRO and \replayRN. When a
replaying process detects \emph{all} the four different ground-truth concurrency-bugs, we record the total execution time (in \emph{minutes}). Table~\ref{tbl:BugsMT-tte} shows the results.

In Table~\ref{tbl:BugsMT-tte}, compared to \replayRO, we 
can observe that \replayRN reduces the average time-to-exposure from 66.5 minutes to 34.1 minutes. This fact means, for example, given a tighter replay time budget (say, 60 minutes), \replayRO has a high chance to miss some of the four concurrency-bugs.
Moreover, \replayRN is more stable since the timing variance is much smaller than that of 
\replayRO (91.0 vs. 959.2). This result implicates that, in Table~\ref{tbl:eval_BugsMT}, for the concurrency-bug revealing capability of \mafl, the \replayRN's result in \projfont{gm-cnvt} is likely to be much better than \replayRO's. 

The evaluation of time-to-exposure suggests that, given a set of seeds, \replayRN is prone to expose concurrency-bugs
faster and more stable. Since \replayRN is closely relevant to schedule-intervention
instrumentation (\S\ref{sec:instrument_schedule}) and repeated execution (\S\ref{sec:cal_exec}),
this also indicates that these strategies are helpful for concurrency-bug revealing.

\begin{table}[t]
\caption{Time-to-exposure of \projfont{gm-cnvt}'s concurrency-bugs during six replays with patterns \replayRO and \replayRN.}
\label{tbl:BugsMT-tte}
\centering
\begingroup
\setlength{\tabcolsep}{5pt} 
\renewcommand{\arraystretch}{0.9} 
\footnotesize
\footnotesize
\begin{tabular}{crrrrrrrr}
\toprule
 & \#1 & \#2 & \#3 & \#4 & \#5 & \#6 & Avg & Variance \\
 \midrule
\replayRO  & 55.3 & 92.1 & 21.8 & 93.7 & 101.5 & 34.7 & 66.5 & 959.2 \\ 
\replayRN  & 33.4 & 52.2 & 33.5 & 37.6 & 24.7 & 23.3 & 34.1 & 91.0 \\ 
\bottomrule
\end{tabular}
\endgroup
\end{table}
 
\subsubsection{Time Budget During Replaying}\label{sec:app-replay-budget}
We chose two hours (2h) as the time budget in the reply phase during evaluation.
Unlike the fuzzing phase, which aims to generate \emph{new seed} files that exercise multithreading context, the replay phase runs the target program against \emph{existing seeds} (generated during fuzzing). Therefore, the criterion is to 1) minimize the time for replay; 2) ensure that replay phase traverses all the generated seeds.
For projects with less generated (multithreading-relevant) seeds (e.g., \testsMT=126 for \projfont{pbzip2-c} when applying \muzz), traversing the seeds (with both \replayRO and \replayRN) \emph{once} are quite fast; however for projects with more generated seeds (e.g., \testsMT=13774 for \projfont{gm-cnvt} when applying \muzz), this requires more time. To make the evaluation fair, we use the {fixed} time budget for all the 12 benchmarks, where seeds in projects like \projfont{pbzip2-c} will be traversed repeatedly until timeout. During the evaluation, we found 2h to be moderate since it can traverse all the generated \emph{multithreading-relevant seeds} at least once for all the projects.

Less time budget, e.g., 1h, may make the replay phase to miss certain generated seeds triggering concurrency violation conditions. In fact, from Table~\ref{tbl:BugsMT-tte}, we see that time-to-exposure for the concurrency-bugs may take 101.5 minutes. Meanwhile, more time budget, e.g., 4h, might be a waste of time for the exercised 12 benchmarks. In fact, in a case study for \projfont{gm-cnvt}, when time budget is 4h, despite that {\replayCN} is nearly doubled, the number of revealed \BugsMT (i.e., {\replayCB}) is still the same as the results in Table~\ref{tbl:eval_BugsMT}, regardless of {\replayRO} or \replayRN.

\subsubsection{Statistical Evaluation Results}\label{sec:app-more-stats}

Specific to the nature of multithreaded programs and our evaluation strategy to determine seeds' relevance with multithreading, we decide not to provide some commonly-used statistical results~\cite{ccs18_eval_fuzzing}.

First, it is \emph{unfair} to track \emph{coverage over time} when comparing \muzz, \mafl with AFL or \mopt due to the different meanings of ``coverage''. In fact, owing to coverage-oriented instrumentation (in \muzz) and threading-context instrumentation (in \muzz and \mafl), \muzz and \mafl cover more execution states (corresponding to \testsALL), therefore naturally preserve more seeds. That is also the reason that in \S\ref{sec:rq_seeds} the values of \testsMT and \testsRatio are more important than \testsALL. 

Second, we \emph{cannot} compare the \emph{multithreading-relevant paths over time} among \muzz, \mafl, AFL, and \mopt. This reason is simple: we resort to a separate procedure \emph{after fuzzing} to determine whether it covers thread-forking routines. We have to do so since AFL and \mopt do not provide a builtin solution to discovering seeds' relevance with multithreading. Consequently, we cannot plot multithreading-relevant crashing states over time.

Third, despite that the \emph{statistical variance} is important, it is not easy to be calculated comprehensively. During evaluation, to reduce the variance among individuals, we apply an \emph{ensemble strategy} by sharing seeds among the six runs, for each of the specific fuzzers~\cite{afl_detail}. However, for multithreaded target programs, another variance comes from the \emph{thread scheduling} for different threads (in our experiments, four working threads were specified). \muzz and \mafl have the schedule-intervention instrumentation to help diversify the effects, while it is absent in AFL and \mopt. In fact, from the case studies in ~\S\ref{sec:app-schedule-intervention}, we envision that the variance may be huge for different machines under different workloads. Due to this, providing \emph{fair} statistical results w.r.t. the variance may still be impractical. Therefore, we tend to exclude variance metrics and only choose those that exhibit the ``overall results'', i.e., \testsMT, \testsRatio, \crashesMT, \NVulsMT, \replayCN, and \replayCB. Similarly, the case studies or comparisons in ~\S\ref{sec:rq_seeds}, ~\S\ref{sec:rq_Vuls}, ~\S\ref{sec:rq_BugsMT} are all based on ``overall results''.
During the evaluation, we indeed observed that the results of \muzz and \mafl are more stable than those of AFL and \mopt.

\section{Related Work}\label{sec:related}

\subsection{Grey-box Fuzzing Techniques}

The most relevant is the fuzzing techniques on concurrency-vulnerability detection. ConAFL~\cite{ConAFL} is a thread-aware GBF that focuses on user-space multithreaded programs. Much different from \muzz's goal to reveal both {\VulsMT} and \BugsMT, ConAFL only detects a subset of concurrency-bug induced vulnerabilities (\VulsCB) that cause \emph{buffer-overflow}, \emph{double-free}, or \emph{use-after-free}. ConAFL also utilizes heavy thread-aware static and dynamic analyses, making it suffer from scalability issues. The other difference is that \muzz's thread-aware analyses aim to provide runtime feedback to distinguish more execution states in multithreading contexts, to bring more multithreading-relevant seeds; meanwhile, ConAFL relies on the discovery of sensitive concurrency operations to capture pairs that may introduce the aforementioned three kinds of vulnerabilities. Further, since the static and dynamic analyses aim to capture and intervene ``sensitive concurrency operation pairs'', ConAFL suffers from the scalability issue. In fact, the biggest binary it evaluated was 196K (\projfont{bzip2smp}), while \muzz can handle programs scaling to 19.4M (\projfont{im-cnvt}). In the evaluation, we did not evaluate ConAFL --- the GitHub version of ConAFL (\url{https://github.com/Lawliar/ConAFL}) does not work since its static analysis is not publicly available and it is not trivial to implement that technique ourselves; further, we have not obtained the runnable tool after we requested from the authors.
 RAZZER~\cite{razzer} utilizes a customized hypervisor to control thread-interleaving deterministically to trigger data races in Linux kernel. It is a \emph{kernel fuzzer} that cannot reveal multithreading-relevant bugs in \emph{user-space} programs. As a matter of fact, the proof-of-crashes are essentially \emph{sequences of system calls} that could trigger race conditions, and the fix of the detected vulnerabilities requires patches to the \emph{kernel} code. Consequently, the guidance of fuzzing is also different. RAZZER spots the over-approximated racing segments and tames non-deterministic behavior of the kernel such that it can deterministically trigger a race. While \muzz's solution is to distinguish more thread-interleaving states to trap the fuzzing to reveal more multithreading-relevant paths. Practically, it is not easy to effectively sequentialize the thread-interleavings to fuzz the user-space programs~\cite{afl-mt-st}.

Multithreading-relevant bugs are inherently \emph{deep}. To reveal deep bugs in the target programs, some GBFs facilitate other feedback ~\cite{driller,steelix,QSYM,Angora,CollAFL,Wang2020Typestate,Wen2020MemLock,ndss20_TortoiseFuzz}. 
Angora~\cite{Angora} distinguishes different calling context when calculating deputy instruction transitions to keep more valuable seeds. Driller~\cite{driller}, QSYM~\cite{QSYM}, and Savior~\cite{savior} integrate symbolic execution to provide additional coverage information to exercise deeper paths. \muzz inspires from these techniques in that it provides more feedback for multithreading context with stratified coverage-oriented and thread-context instrumentations, as well as schedule-intervention instrumentation. 
Other fuzzing techniques utilize the domain knowledge of the target program to generate more effective seeds~\cite{skyfire,superion,smart_gbf}. Skyfire~\cite{skyfire} and Superion~\cite{superion} provide customized seed generation and mutation strategies on the programs that feed grammar-based inputs. SGF~\cite{smart_gbf} relies on the specifications of the structured input to improve seed quality. These techniques are orthogonal to \muzz and can be integrated into \emph{seed mutation} (c.f. \PfuzzS in Figure~\ref{fig:overview}).

\subsection{Static Concurrency-bug Prediction}
Static concurrency-bug (\BugsMT) predictors aim to approximate the runtime behaviors of a concurrent program without actual execution. Several static approaches have been proposed for analyzing Pthread and Java  programs
~\cite{pratikakis2006locksmith,Vojdani2009,SuiDX16}.
LOCKSMITH~\cite{pratikakis2006locksmith} uses existential types to correlate locks and data in dynamic heap structures for race detection. Goblint~\cite{Vojdani2009} relies on a thread-modular constant propagation and points-to analysis for detecting concurrent bugs by considering conditional locking schemes.
\cite{voung2007relay} scales its detection to large codebases by sacrificing soundness and suppressing false alarms using heuristic filters.
FSAM~\cite{SuiDX16,Sui:2016:SVF} proposes a sparse flow-sensitive pointer analysis for C/C++ programs using context-sensitive thread-interleaving analysis.
Currently, \muzz relies on flow- and context-insensitive results of FSAM for thread-aware instrumentations.
We are seeking solutions to integrating other bug prediction techniques to further improve \muzz's effectiveness.

\subsection{Dynamic Analysis on Concurrency-bugs}
There are a large number of dynamic analyses on concurrency-bugs. They can be divided into two categories: modeling concurrency-bugs and strategies to trigger these bugs.

The techniques in the first category
~\cite{pldi09_fasttrack,lockset_SavageABNS97,kcc:tsan,YuNPP12}
typically monitor the memory and synchronization events~\cite{mtbugs_survey}. The two fundamentals are \emph{happens-before model}~\cite{pldi09_fasttrack} and \emph{lockset model}~\cite{lockset_SavageABNS97}. Happens-before model reports a race condition when two threads read/write a shared memory arena in a causally unordered way, while at least one of the threads write this arena. Lockset model conservatively considers a potential race if two threads read/write a shared memory arena without locking. Modern detectors such as \ts~\cite{kcc:tsan}, Helgrind~\cite{helgrind} usually apply a hybrid strategy to combine these two models. 
\muzz \emph{does not} aim to improve existing concurrency violation models; instead, it relies on these models to detect concurrency-bugs with our fuzzer-generated seeds.

The second category of dynamic analyses focuses on how to trigger concurrency violation conditions. This includes random testings that mimic non-deterministic program executions~\cite{JoshiPSN09,ParkS08,CaiC12}, regression testings~\cite{TerragniCZ15,YuHW18} that target interleavings from code changes, model checking~\cite{FlanaganG05,ZaksJ08,yang2008inspect} and hybrid constraint solving~\cite{pldi14_maxmodel,Huang15,icse18_ufo} approaches that systematically check or execute possible thread schedules, heuristically avoid fruitless executions~\cite{ASPLOS11conseq,GuoKWYG15,GuoKW16,erlang-cbug}, or utilizing multicore to accelerate bug detection~\cite{DBLP:conf/pldi/NagarakatteBMM12}. 
Our work differs from all the above, as our focus is \emph{not} to test schedules with a given seed file, but to generate {seed files} that execute multithreading-relevant paths. In particular, our goal of schedule-intervention instrumentation is to diversify the actual schedules to help provide feedback during fuzzing.

\section{Conclusion}
This paper presented \muzz, a novel technique that empowers thread-aware seed generation to GBFs for fuzzing multithreaded programs. Our approach performs three novel instrumentations that can distinguish execution states introduced by thread-interleavings. Based on the feedback provided by these instrumentations, \muzz optimizes the dynamic strategies to stress different kinds of multithreading context. 
Experiments on 12 real-world programs demonstrate that \muzz outperforms other grey-box fuzzers such as AFL and \mopt in generating valuable seeds, detecting concurrency-vulnerabilities, as well as revealing concurrency-bugs.
 \section*{Acknowledgement}
This research was supported (in part) by the National Research
Foundation, Prime Ministers Office, Singapore under its National
Cybersecurity R\&D Program (Award No. NRF2018NCR-NCR005-0001), National Satellite of Excellence in Trustworthy Software
System (Award No. NRF2018NCR-NSOE003-0001), and NRF Investigatorship (Award No. NRFI06-2020-0022) administered by
the National Cybersecurity R\&D Directorate. The research of Dr
Xue is supported by CAS Pioneer Hundred Talents Program. 


\end{document}